\documentclass[twoside,onecolumn,9pt]{article}
\usepackage{extsizes}
\usepackage[super,sort&compress,comma]{natbib} 
\usepackage[version=3]{mhchem}
\usepackage[left=1.5cm, right=1.5cm, top=1.785cm, bottom=2.0cm]{geometry}
\usepackage{sectsty}
\usepackage{graphicx} 
\usepackage{lastpage}
\usepackage[format=plain,justification=justified,singlelinecheck=false,font={stretch=1.125,small,sf},labelfont=bf,labelsep=space]{caption}
\usepackage{float}
\usepackage{fancyhdr}
\usepackage{fnpos}
\usepackage[english]{babel}
\addto{\captionsenglish}{%
  
}
\usepackage{array}
\usepackage{droidsans}
\usepackage{charter}
\usepackage[T1]{fontenc}
\usepackage[usenames,dvipsnames]{xcolor}
\usepackage{setspace}
\usepackage[compact]{titlesec}
\usepackage{hyperref}
\usepackage{epstopdf}%This line makes .eps figures into .pdf - please comment out if not required.

\usepackage[utf8]{inputenc}
\usepackage{graphicx,subfig,amsmath,bbm}
\usepackage{bm}
\usepackage{amsmath}
\usepackage{amssymb}
\usepackage{amsfonts}
\usepackage{alphalph}

\newcommand{\bra}[1]{\left< \smash{ #1 } \right|}
\newcommand{\ket}[1]{\left| \smash{ #1 } \right>}

\newcommand{\mvec}[1]{\mathbf{ #1 }}

\newcommand{\abs}[1]{\left| #1 \right|}

\newcommand{\expo}[2][]{e^{#1 i \, #2}}
\newcommand{\polvec}[1]{{\bm{ \epsilon}}_{#1}}

\newcommand{\Aqu}[3][]{\hat{\mvec{A}}_{\textsc{#1}}^{#2}(#3)}

\newcommand{\ZH}{Z}
\newcommand{\argline}[8]{(\smash{ \mvec{k}^{#1}_{#2}, {\omega}^{#3}_{#4}, \mvec{k}^{#5}_{#6}, {\omega}^{#7}_{#8}})}
\newcommand{\PProp}[1]{\mvec{P}_{#1}}

\definecolor{cream}{RGB}{222,217,201}

\begin{document}

\pagestyle{fancy}
\thispagestyle{plain}
\fancypagestyle{plain}{
%%%HEADER%%%
\renewcommand{\headrulewidth}{0pt}
}
%%%END OF HEADER%%%

%%%PAGE SETUP - Please do not change any commands within this section%%%
\makeFNbottom
\makeatletter
\renewcommand\LARGE{\@setfontsize\LARGE{15pt}{17}}
\renewcommand\Large{\@setfontsize\Large{12pt}{14}}
\renewcommand\large{\@setfontsize\large{10pt}{12}}
\renewcommand\footnotesize{\@setfontsize\footnotesize{7pt}{10}}
\makeatother

\renewcommand{\thefootnote}{\fnsymbol{footnote}}
\renewcommand\footnoterule{\vspace*{1pt}% 
\color{cream}\hrule width 3.5in height 0.4pt \color{black}\vspace*{5pt}} 
\setcounter{secnumdepth}{5}

\makeatletter 
\renewcommand\@biblabel[1]{#1}            
\renewcommand\@makefntext[1]% 
{\noindent\makebox[0pt][r]{\@thefnmark\,}#1}
\makeatother 
\renewcommand{\figurename}{\small{Fig.}~}
\sectionfont{\sffamily\Large}
\subsectionfont{\normalsize}
\subsubsectionfont{\bf}
\setstretch{1.125} %In particular, please do not alter this line.
\setlength{\skip\footins}{0.8cm}
\setlength{\footnotesep}{0.25cm}
\setlength{\jot}{10pt}
\titlespacing*{\section}{0pt}{4pt}{4pt}
\titlespacing*{\subsection}{0pt}{15pt}{1pt}
%%%END OF PAGE SETUP%%%

%%%FOOTER%%%
\fancyfoot{}
\fancyfoot[LO,RE]{\vspace{-7.1pt}\includegraphics[height=9pt]{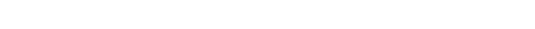}}
\fancyfoot[CO]{\vspace{-7.1pt}\hspace{13.2cm}\includegraphics{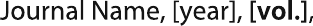}}
\fancyfoot[CE]{\vspace{-7.2pt}\hspace{-14.2cm}\includegraphics{head_foot/RF}}
\fancyfoot[RO]{\footnotesize{\sffamily{1--\pageref{LastPage} ~\textbar  \hspace{2pt}\thepage}}}
\fancyfoot[LE]{\footnotesize{\sffamily{\thepage~\textbar\hspace{3.45cm} 1--\pageref{LastPage}}}}
\fancyhead{}
\renewcommand{\headrulewidth}{0pt} 
\renewcommand{\footrulewidth}{0pt}
\setlength{\arrayrulewidth}{1pt}
\setlength{\columnsep}{6.5mm}
\setlength\bibsep{1pt}
%%%END OF FOOTER%%%

%%%FIGURE SETUP - please do not change any commands within this section%%%
\makeatletter 
\newlength{\figrulesep} 
\setlength{\figrulesep}{0.5\textfloatsep} 

\newcommand{\topfigrule}{\vspace*{-1pt}% 
\noindent{\color{cream}\rule[-\figrulesep]{\columnwidth}{1.5pt}} }

\newcommand{\botfigrule}{\vspace*{-2pt}% 
\noindent{\color{cream}\rule[\figrulesep]{\columnwidth}{1.5pt}} }

\newcommand{\dblfigrule}{\vspace*{-1pt}% 
\noindent{\color{cream}\rule[-\figrulesep]{\textwidth}{1.5pt}} }

\makeatother
%%%END OF FIGURE SETUP%%%

%%%TITLE, AUTHORS AND ABSTRACT%%%
\onecolumn
  \begin{@twocolumnfalse}
{\includegraphics[height=30pt]{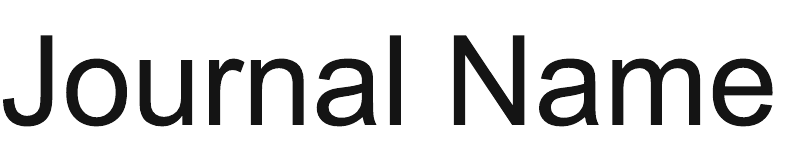}\hfill\raisebox{0pt}[0pt][0pt]{\includegraphics[height=55pt]{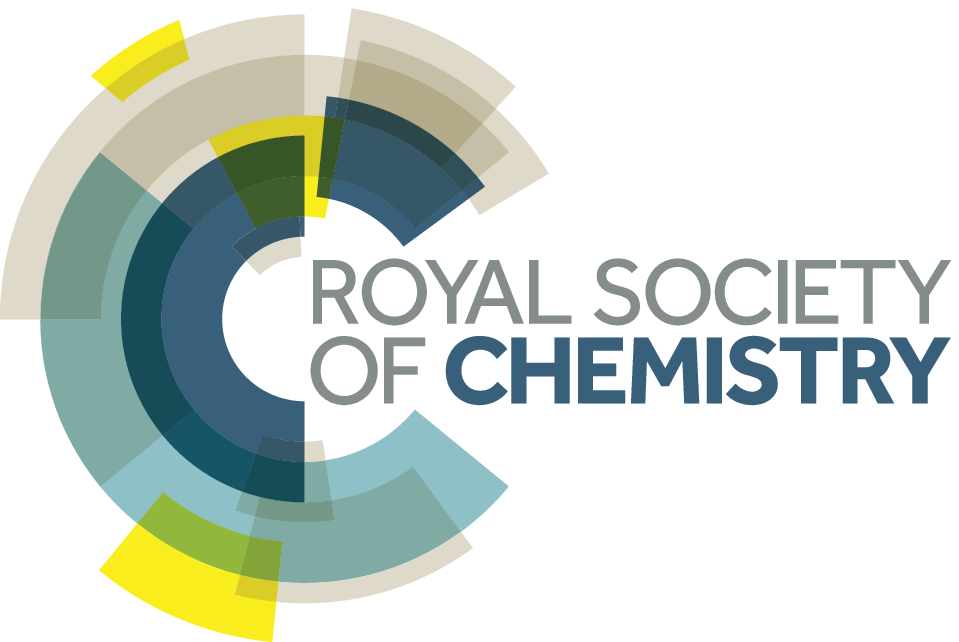}}\\[1ex]
\includegraphics[width=18.5cm]{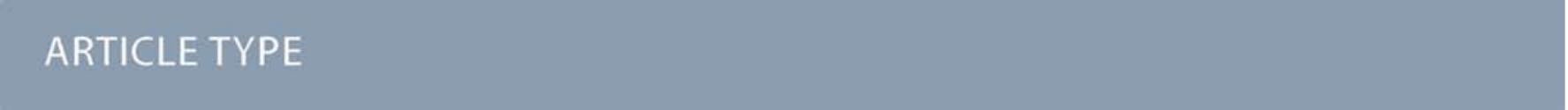}}\par
\vspace{1em}
\sffamily
\begin{tabular}{m{4.5cm} p{12.5cm} }

\includegraphics{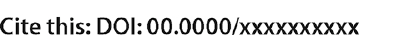} & \noindent\LARGE{\textbf{Towards novel probes for valence charges via x-ray optical wave mixing}} \\%Article title goes here instead of the text "This is the title"
\vspace{0.3cm} & \vspace{0.3cm} \\

 & \noindent\large{Christina Boemer\textit{$^{a\ddag}$}, Dietrich Krebs\textit{$^{b, c\ddag}$}, Andrei Benediktovitch\textit{$^{a}$}, Emanuele Rossi\textit{$^{b,d}$}, Simo Huotari\textit{$^{e}$}, and Nina Rohringer\textit{$^{a,b,c,d}$}} \\%Author names go here instead of "Full name", etc.

%!!!! Affiliations to be set in footnotes below - put letters manually

\includegraphics{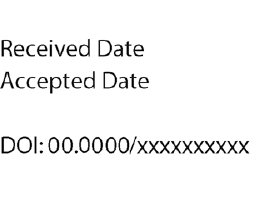} & \noindent\normalsize{We present a combined theoretical and experimental study of x-ray optical wave mixing. This class of nonlinear phenomena combines the strengths of spectroscopic techniques from the optical domain, with the high-resolution capabilities of x-rays. In particular, the spectroscopic sensitivity of these phenomena can be exploited to selectively probe valence dynamics. Specifically, we focus on the effect of x-ray parametric down-conversion. We present a theoretical description of the process, from which we deduce the observable nonlinear response of valence charges. Subsequently, we simulate scattering patterns for realistic conditions and identify characteristic signatures of the nonlinear conversion. For the observation of this signature, we present a dedicated experimental setup and results of a detailed investigation. However, we do not find evidence of the nonlinear effect. This finding stands in strong contradiction to previous claims of proof-of-principle demonstrations. Nevertheless, we are optimistic to employ related x-ray optical wave mixing processes on the basis of the methods presented here for probing valence dynamics in the future.} \\%The abstract goes here instead of the text "The abstract should be..."

\end{tabular}

 \end{@twocolumnfalse} \vspace{0.6cm}

%%%END OF TITLE, AUTHORS AND ABSTRACT%%%

%%%FONT SETUP - please do not change any commands within this section
\renewcommand*\rmdefault{bch}\normalfont\upshape
\rmfamily
\section*{}
\vspace{-1cm}

%%%FOOTNOTES%%%
\footnotetext{\textit{$^{a}$~Deutsches Elektronen Synchrotron DESY, Notkestrasse 85, 22607 Hamburg, Germany.}}
\footnotetext{\textit{$^{b}$~Department of Physics, Universit\"at Hamburg, Jungiusstrasse 9, 20355 Hamburg, Germany.}}
\footnotetext{\textit{$^{c}$~Max Planck School of Photonics, Friedrich-Schiller University of Jena, Albert-Einstein-Str. 6, 07745 Jena, Germany.}}
\footnotetext{\textit{$^{d}$~The Hamburg Centre for Ultrafast Imaging, Luruper Chaussee 149, 22761 Hamburg, Germany.}}
\footnotetext{\textit{$^{e}$~Department of Physics, University of Helsinki, Gustaf H\"allstr\"omin katu 2, 00560 Helsinki, Finnland.}}

%Please use \dag to cite the ESI in the main text of the article.
%If you article does not have ESI please remove the the \dag symbol from the title and the footnotetext below.
%\footnotetext{\dag~Electronic Supplementary Information (ESI) available: [details of any supplementary information available should be included here]. See DOI: 00.0000/00000000.}
%additional addresses can be cited as above using the lower-case letters, c, d, e... If all authors are from the same address, no letter is required

\footnotetext{\ddag~These authors contributed equally to this work. Correspondence via E-mail: christina.boemer@desy.de, dietrich.krebs@desy.de, or nina.rohringer@desy.de}

%%%END OF FOOTNOTES%%%

%alphalph workaround for footnotes
\makeatletter
\newalphalph{\fnsymbolwrap}[wrap]{\@fnsymbol}{}
\makeatother\renewcommand*{\thefootnote}{%
   \fnsymbolwrap{\value{footnote}}%
 }

%%%MAIN TEXT%%%%
\section{Introduction}
The functionality of materials is encoded in their electronic structure.
More specifically, it is the structure of the valence electrons and their response behaviour that determine the properties of a material.
In order to investigate these, a plethora of methods has been developed, many of which can be counted to the domain of optical spectroscopy in the visible and ultraviolet spectral domain.
Starting from linear spectroscopic techniques, which probe electronic transitions and associated energies, nonlinear spectroscopic techniques have been developed, which probe more complex response functions and yield a wealth of multidimensional information \cite{2002Shen_BOOK-Nonlinear_Optics,1999Mukamel-BOOK}.
All of these allow for valuable information on the electronic structure of a sample, yet---in a spatial sense---this structure still remains unresolved by optical techniques. Involving probes of x-ray wavelengths provides a clear path to solve this problem. As highly brilliant x-ray sources become operational alongside sophisticated optical lasers,
techniques are developed that combine their respective strengths, i.e., using the valence specificity of optical excitations and the spatial resolving power of x-rays. A prominent example of such a combination is fs time-resolved x-ray diffraction, following laser-induced redistribution of the electron density: By populating an excited electronic state through laser excitation, x-ray diffraction with fs pulses probes the adiabatic electronic redistribution due to nuclear motion on the excited nuclear potential manifolds. The extension of these methods to follow electronic wave packets on the attosecond time scale is not straight forward, both for conceptual and technical reasons \cite{2012Dixit-PNAS,2017Kowaleski-nonadiabatic_XRD,2019Simmermacher-coherence}. In these concepts an optical pump pulse is envisioned to prepare a valence-electron wave packet and a subsequent x-ray pulse probes its dynamics by time-resolved x-ray scattering. In addition to these optical pump x-ray probe methods, there have been proposals to transfer higher order wave mixing concepts from nonlinear optics to the x-ray domain (cf. Ref. \cite{2018Young-XrayRoadmap} and references therein). Such methods could be used to probe complex dynamical processes with spatial resolution of the corresponding nonlinear response functions.\\
\\
In the present work, we want to focus on a comparatively simple process from the broad range of potential nonlinear probes.
Specifically, we will discuss x-ray-optical three-wave mixing in the form of (spontaneous) x-ray parametric down-conversion (XPDC).
This process has attracted some attention recently, as down-conversion of a hard x-ray photon ("pump") into a pair of x-ray ("signal") and optical ("idler") photon was claimed for the first time\cite{schori2017parametric}.
Subsequent publications report on the spectral sensitivity of XPDC both for single particle and collective valence excitations in crystalline solids\cite{borodin2017high,sofer2019observation,2019Borodin-SharonPlasmon}.
If confirmed, these would mark highly desirable features of a nonlinear probe and thus warrant in-depth analysis.
At the same time, however, our first independent investigation of XPDC has raised serious concerns with the aforementionned reports\cite{boemer2020x}.\\
Now, we put our discussion of XPDC on a broader basis---providing both a theoretical description of the process and novel experimental results.
Within our theoretical framework, we, first of all, identify the nonlinear response function that is accessible via XPDC. In order to visualize the probed transitions, we present a spatially resolved example for the case of a diamond sample (Sec.~\ref{ssec:nl-response}).
Moreover, we derive an expression for realistic scattering probabilities and predict the ensuing scattering patterns (Sec.~\ref{ssec:nl-response}).
Complementing the theoretical approach, we discuss the construction and characterization of an experimental setup dedicated to the search for XPDC (Sec.~\ref{ssec:setup_cal}).
While we can report on significant improvements in resolving power compared to all previous investigations, we do \emph{not} observe XPDC (Sec.~\ref{ssec:exp_results}).
The upper bound, which we deduce for the nonlinear conversion efficiency, is compatible with our theoretical predictions, yet both obviously disagree with the aforementionned claims on XPDC (Sec.~\ref{sec:controversy}).
Finally, we discuss ways to combat the low count rates of x-ray-optical wave mixing and use our improved crystal optics to map out other forms of three-wave mixing.
Thereby, we conclude that nonlinear x-ray probes of valence dynamics are nevertheless within reach---albeit not in terms of XPDC (Sec.~\ref{sec:outlook}).

%\section{footnotes}
%For Dietrich:
%For footnotes in the main text of the article please number the footnotes to avoid duplicate symbols. \textit{e.g.}\ \texttt{\textbackslash footnote[num]\{your text\}}. The corresponding author $\ast$ counts as footnote 1, ESI as footnote 2, \textit{e.g.}\ if there is no ESI, please start at [num]=[2], if ESI is cited in the title please start at [num]=[3] \textit{etc.} Please also cite the ESI within the main body of the text using \dag.

\section{Theoretical description of XPDC}
In order to advance our understanding of XPDC, we have developed a first-principles description of the process.
This serves as a guide to the expectable scattering signal and also provides a basis for interpreting the nonlinear response.
Conversely, our experimental study will provide a benchmark for the theory's validity.
Our approach to XPDC is based on a more general description of parametric x-ray optical wave mixing processes \cite{202XKrebs} that we have formulated within the framework of non-relativistic QED \cite{1998CraigThirunamachandran-BOOK}. 
In the following, we present our central observable, which we derived by assuming a scattering perspective for non-resonant x-ray interactions. It is adapted to the case of XPDC and reads\footnote[2]{Note that throughout the theoretical discussion, the system of atomic units will be used---unless otherwise indicated. This entails the simplifying prescription \mbox{$\hbar = m_\text{e} = \abs{e} = 1$}, where we set the reduced Planck constant, the mass and the charge of an electron, respectively, equal to unity.
Furthermore, the speed of light in vacuo is referenced to $c = 1/\alpha$, with $\alpha$ denoting the fine structure constant.
In addition, we adopt the summation convention that doubly occurring greek indices are summed over---spanning the range of $1,2,3$ for three-component vectors.}:
\begin{align}
  \frac{d P_{\textsc{xpdc}}(\mvec{k}_s)}{d\Omega_s d\omega_s}
  &= 
  \frac{\alpha^7 \, \omega_s \, (\polvec{s})_\sigma  (\polvec{s}^*)_\rho}{(2\pi)^{16}} \,
  \int \!\! d\omega_i \, d\omega_i^\prime \, d\omega_p \, d\omega_p^\prime \!\!\! \int \!\! d^3k_i \,  d^3k_i^\prime \, d^3k_p \, d^3k_p^\prime \,  
  \delta(\omega_s + \omega_i - \omega_p) \, 
  \delta(\omega_s + \omega_i^\prime - \omega_p^\prime)  \nonumber \\
  &\times
   [{\ZH}^{(1)}_{\textsc{pump}} \argline{\prime}{p}{\prime}{p}{}{p}{}{p}]_{\sigma\rho}  \,
   [\bar{C}^{(1)}_{\textsc{vac}} (\smash{ -\mvec{k}^{\prime}_{i}, {-\omega}^{\prime}_{i}, -\mvec{k}_{i}, {-\omega}_{i}})]_{\nu\mu} ~
  \left[ \mvec{K}(\mvec{k}_i,\mvec{k}_s-\mvec{k}_p,-\omega_i) \right]_\mu
  \left[ \mvec{K}(\mvec{k}_i^\prime,\mvec{k}_s-\mvec{k}_p^\prime,-\omega_i^\prime) \right]^*_\nu  
  .
  \label{eq:observable_kw}
\end{align}
This gives the double-differential probability to find a down-converted x-ray photon with polarization $\polvec{s}$ scattered into an element of solid angle $d\Omega_s$ and energy range $d\omega_s$ around the wave vector $\mvec{k}_s$.
The signal photon is converted from the x-ray pump pulse---characterized by its first order correlation function ${\ZH}^{(1)}_{\textsc{pump}}$---while simultaneously an optical idler photon is created from the vacuum. The vacuum fluctuations governing this creation process are described by a similar correlator, viz., $\bar{C}^{(1)}_{\textsc{vac}}$.
The overall conversion efficiency and spatial dependence of XPDC is determined by the nonlinear response function of the material $\mvec{K}$. 
This captures the electronic excitation dynamics during the wave mixing process and will enable the visualization of involved transitions, if (experimentally) resolved (see discussion in the following section).
Finally, it is noteworthy about Eq.~(\ref{eq:observable_kw}) that the parametric nature of the conversion process manifests itself in terms of the two energy conserving $\delta$-functions. 
For further details on the derivation of Eq.~(\ref{eq:observable_kw}), see Ref.~\cite{202XKrebs}.
 
\subsection{Probing nonlinear response}
\label{ssec:nl-response}
\begin{figure}[!ht]
   \centering
   %\subfloat[]{\includegraphics[width=.49\textwidth]{2020-11-29_abssquaredR_sym_in_110_plane.png}}\,\,
   \includegraphics[width=.49\textwidth]{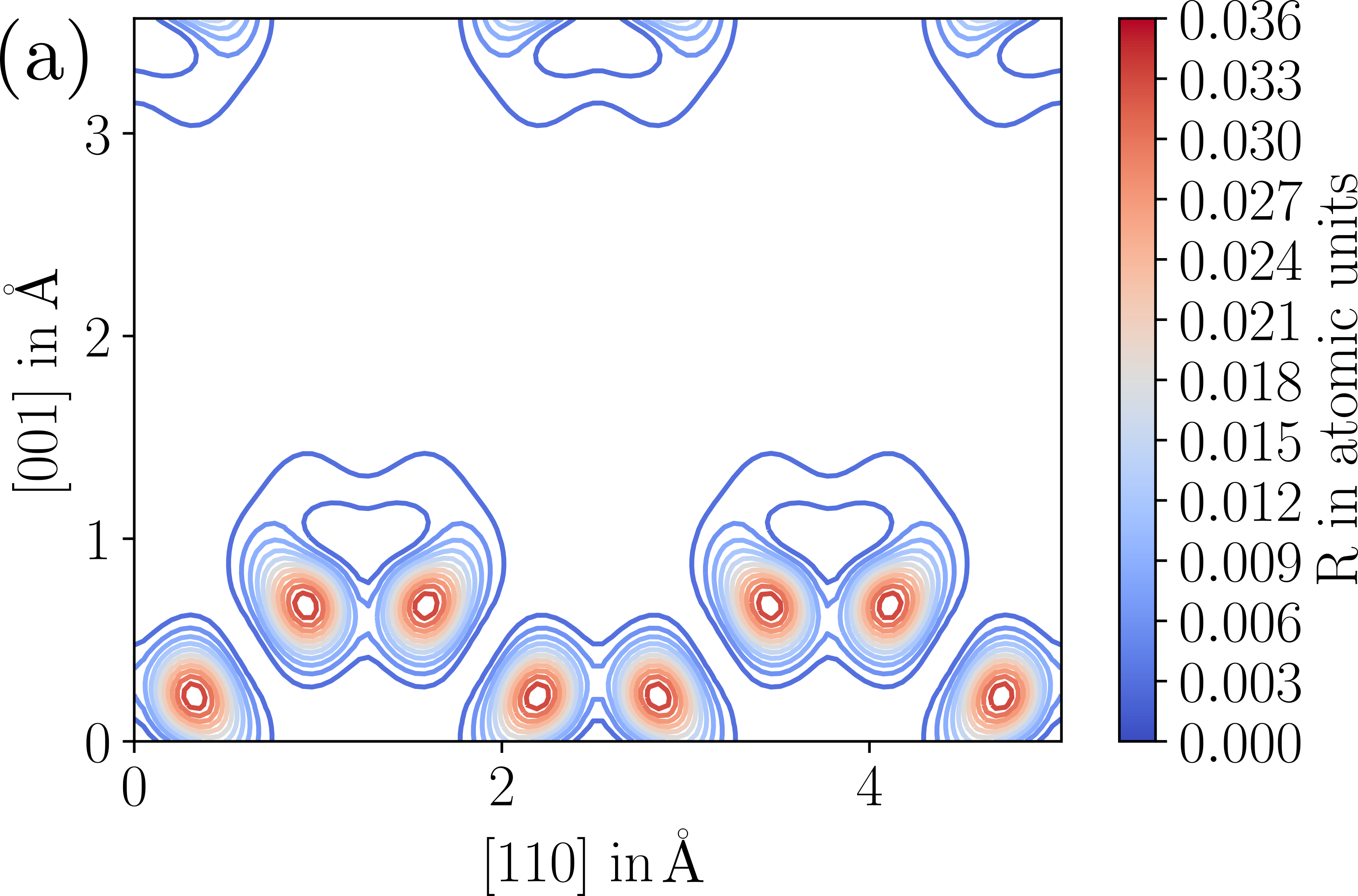}\,\,
   %\subfloat[]{\includegraphics[width=.49\textwidth]{2020-11-29_rho_sym_in_110_plane.png}}\\
   \includegraphics[width=.49\textwidth]{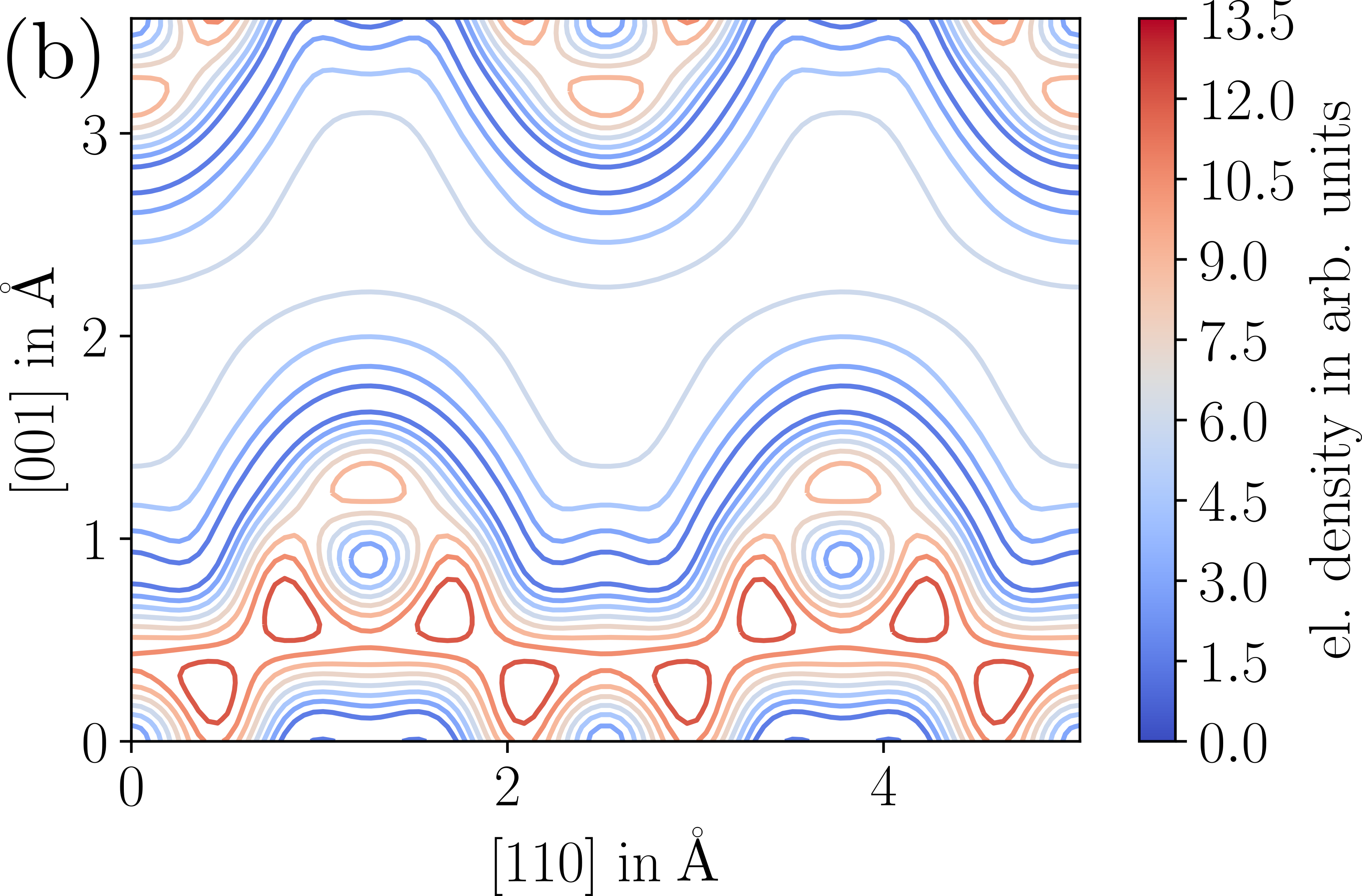}\\
   \caption{(a) The nonlinear response of diamond is localized in two maxima along each bond. This reflects the transition from the bonding ground state into excited states of dominantly anti-bonding character. The 2D cut of $\abs{\smash{\mvec{R}_{\diamond}(\mvec{x},\omega)}}^2$
   is taken in the (110)-plane of the conventional (cubic) unit cell at an excitation energy of $\omega = 0.184 \text{ a.u. } (\sim 5.0 \text{ eV})$.
(b) For reference, a cut through the unperturbed valence electron density $\rho_{\textsc{val}}(\mvec{x})$ is plotted in the same plane.}
   \label{fig:response2D}
\end{figure}

The central component of Eq.~(\ref{eq:observable_kw}) is the nonlinear response function $\mvec{K}(\mvec{k}_1,\mvec{k}_2,\omega)$, which embodies the electronic dynamics during the wave mixing process. It is our ultimate aim to obtain access to this quantity from measurements of XPDC. This would---ideally---yield spatio-temporal information on the involved valence transitions via the inverse Fourier transforms
\begin{align}
  \PProp{} (\mvec{y},t_2, \, \mvec{x},t_1) 
  &=
  \frac{1}{(2\pi)^7} ~ \int \!\! d^3k_1  \, d^3k_2  \int \!\! d\omega
  ~ \expo{(\mvec{k}_1\cdot\mvec{y} + \mvec{k}_2\cdot\mvec{x})} ~ \expo[-]{\omega(t_1 - t_2)} ~
  \mvec{K}(\mvec{k}_1,\mvec{k}_2,\omega) 
  .
  \end{align}
In the following, we will discuss the nature of this real-space correlator $\PProp{} (\mvec{y},t_2, \, \mvec{x},t_1)$ and simplify it to a point, where we can visualize its spatial dependence. The explicit form of this response function reads
%%%
\begin{align}
  \PProp{} (\mvec{y},t_2, \, \mvec{x},t_1) 
  &=
  \bra{I} \, \hat{T}
  \left[\hat{\mvec{p}}(\mvec{y},t_2) \, \hat{n}(\mvec{x},t_1)\right] \,
  \ket{I} 
  ,
  \label{eq:PProp}
\end{align}
where $\ket{I}$ labels the $N$-electron ground state of the sample, while $\hat{\mvec{p}}(\mvec{y},t_2)$ and $\hat{n}(\mvec{x},t_1)$ are the Heisenberg operators of electronic momentum density and electronic particle density, respectively. The symbol $\hat{T}[...]$ denotes the time-ordering operator for all bracketed arguments. The above object is essentially a polarization propagator \cite{1984Oddershede-PolProp} of general form. Both operators connect the ground state on either side of the expression to an intermediate manifold of excited states. In doing so, the momentum density is associated with the (long-wavelength) optical interaction of the XPDC process, while the density operator $\hat{n}(\mvec{x},t_1)$ corresponds to the x-ray interactions
\footnote[3]{It should be noted that this description of parametric down-conversion in the x-ray regime significantly differs from its all-optical counterpart, which would involve three actions of the operator $\hat{\mvec{p}}$ in conjunction with the dipole-approximation \cite{2002Shen_BOOK-Nonlinear_Optics}.}. 
Ultimately, this density operator yields spatial information about the sample. Notably, however, it does not yield a direct image of the electronic density, but probes a ``transition-density'' instead. This observable displays features of the electronic ground state as well as the valence-excited states that partake in the wave mixing process.\\
\\
It is illustrative to visualize the observable part of $\PProp{} (\mvec{y},t_2, \, \mvec{x},t_1)$  for a simple system such as diamond (see Fig.~\ref{fig:response2D}). 
%old Figure here
To this end, we first simplify the overall expression by averaging over the coordinate associated with the optical coupling (i.e., $\mvec{y}$).
Thus, the essential structure of the nonlinear response is captured by
%%%
\begin{align}
  \mvec{R}_{\diamond}(\mvec{x},\omega)
  &=
  \int \!\! d\tau \, \expo[]{\omega \tau} 
  \int_{\diamond} \! d^3y ~ 
  \PProp{\diamond} (\mvec{y},t_2, \, \mvec{x},t_1) 
  =
  \frac{1}{V_\diamond} 
  \sum^{\text{rec.}}_{\mvec{G}} 
  \expo{\mvec{G}\cdot\mvec{x}} ~
  \mvec{K}_{\diamond}(0, \mvec{G},\omega) 
  .
  \label{eq:Rdens}
\end{align}
In writing Eq.~(\ref{eq:Rdens}), we have implemented two further operations.
First, we have imposed periodic boundary conditions on the remaining coordinate ($\mvec{x}$) within a large, fictitious crystal volume ${V_\diamond}$. Under these conditions, the reciprocal representation of the nonlinear response function $\mvec{K}(\mvec{k}_1,\mvec{k}_2,\omega)$ reduces to the simpler $\mvec{K}_{\diamond}(0, \mvec{G},\omega)$, which is discretized to the reciprocal lattice of the crystal \footnote[4]{In order to relate the simplified response function $\mvec{K}_{\diamond}(0, \mvec{G},\omega)$ from the pseudo-infinite crystal back to a realistically finite sample, we employ a window function approach \cite{1992Giacovazzo-BOOK}. The corresponding approximation reads: $\mvec{K}(\mvec{k}_1, \mvec{k}_2,\omega) \approx
  \sum^{\text{rec.}}_{\mvec{G}} \, \tilde{w}(\mvec{k}_1 + \mvec{k}_2 - \mvec{G}) ~  
  {V_\diamond}^{-1} \, \mvec{K}_{\diamond}(\mvec{0}, \mvec{G},\omega)$, where $\tilde{w}$ is the Fourier transformed shape function of the real sample.}. 
Second, we have Fourier transformed the time-dependence $\tau = t_1 - t_2$ to obtain a spectral representation of the nonlinear response.
The resulting $\mvec{R}_{\diamond}(\mvec{x},\omega)$ gives a spatially resolved view of electronic transitions that contribute to the nonlinear response at (transition-) energy $\omega$. Tuning $\omega$ across the manifold of excited states, we can use the spectral selectivity to focus on different contributions to the overall valence dynamics. \\
For our illustrating example, we remain at a single energy \mbox{$\omega = 0.184 \text{ a.u. } (\sim 5.0 \text{ eV})$}, which is below the band gap of diamond. Thus, we are mostly probing the lowest optical transitions of the system. 
We model the electronic structure using Density Functional Theory and evaluate the nonlinear response function $\mvec{K}_{\diamond}(0, \mvec{G},\omega)$ from Kohn-Sham orbitals\footnote[5]{The nonlinear response function is evaluated to first approximation from the Kohn-Sham orbitals of an LDA-DFT calculation performed with the \textsc{abinit} package\cite{2020Gonze-Abinit}. For these calculations, we employ a norm-conserving pseudopotential and a plane-wave basis with energy cut-off at $15 \text{ a.u.}$. The conventional (cubic) unit cell of diamond was fixed to a size of $6.741 \text{ a.u.}$ and the reciprocal Brillouin zone was initially sampled at 28 symmetry adapted k-points. The converged SCF result was subsequently extrapolated to 864 k-points that covered the full Brillouin zone homogeneously. In order to account for the notoriously wrong band-gap energy of LDA-DFT\cite{1985Perdew-bandgaperror}, we shift the orbital energies of all unoccupied states by $\Delta E = 0.062 \text{ a.u. } (\sim 1.7 \text{ eV})$ following the scissor correction from Ref.~\cite{2004Botti-scissor}. In addition, we introduce an imaginary part ($\sim$ decay rate) of $\epsilon = 0.007 \text{ a.u. } (\sim 0.2 \text{ eV})$ to the excited state energies as a means of regularizing the expression; this approach is adopted from earlier work in all-optical response calculations by Benedict et al.\cite{1998Benedict-dielectric}.}.
%Or figure here?
%%
%
Plotting the corresponding $\abs{\smash{\mvec{R}_{\diamond}(\mvec{x},\omega)}}^2$ in the (110)-plane of the conventional (cubic) unit cell of diamond, we can clearly observe that the nonlinear response is localized in two maxima along each bond (cf. Fig.~\ref{fig:response2D} (a)). This pattern partially coincides with the localization of valence electrons in the ground state, which is shown in Fig.~\ref{fig:response2D} (b) in terms of the respective electron density $\rho_{\textsc{val}}(\mvec{x})$. Comparing the two distributions, it is nevertheless obvious that the character of $\mvec{R}_{\diamond}(\mvec{x},\omega)$ is decidedly more antibonding. This reflects the contributions of excited states to the nonlinear response. Effectively, $\mvec{R}_{\diamond}(\mvec{x},\omega)$ monitors the transition from ground to excited states and thus factors in features of the bonding and anti-bonding characters of either. 
Remarkably, such transition patterns are implicitly involved in symmetric (all-optical) polarization propagators as well.
However, they only become experimentally resolvable in nonlinear combination with an x-ray probe.

%Now go for the simulation of scattering pattern
%Need to introduce
%a) vacuum fluctuations
%b) description of x-ray field
% then simplify the expression and show plots

\subsection{Scattering simulations}
\label{ssec:simscat}
\begin{figure}[h!]
\begin{center}
\includegraphics[width=0.95\textwidth]{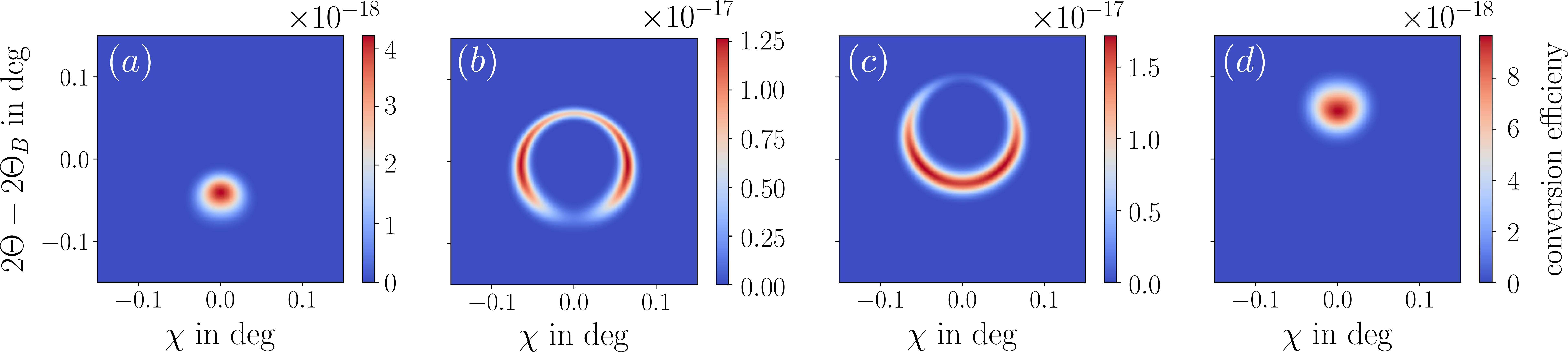}
\caption{XPDC scattering pattern simulated for different rocking angles of the sample, i.e., $\Omega - \Omega_B = -50,~0,~50,~100~\text{mdeg}$ for plots (a) to (d), respectively. The (400) Bragg condition for the fundamental pump energy of $\hbar \omega_{p0} = 11216 \text{ eV}$ is taken as a reference with $\Omega_B = \theta_B$ and the center of each plot set to $2\theta = 2\theta_B$. Each pixel is color coded to show the nonlinear conversion efficiency for its corresponding solid angle ($1 \text{ mdeg} \times 1 \text{ mdeg}$).}
\label{fig:2D-pattern}
\end{center}
\end{figure}

\begin{figure}[!ht]
  \begin{center}  
   \includegraphics[width=.5\textwidth, trim={2cm 0 0 0}]{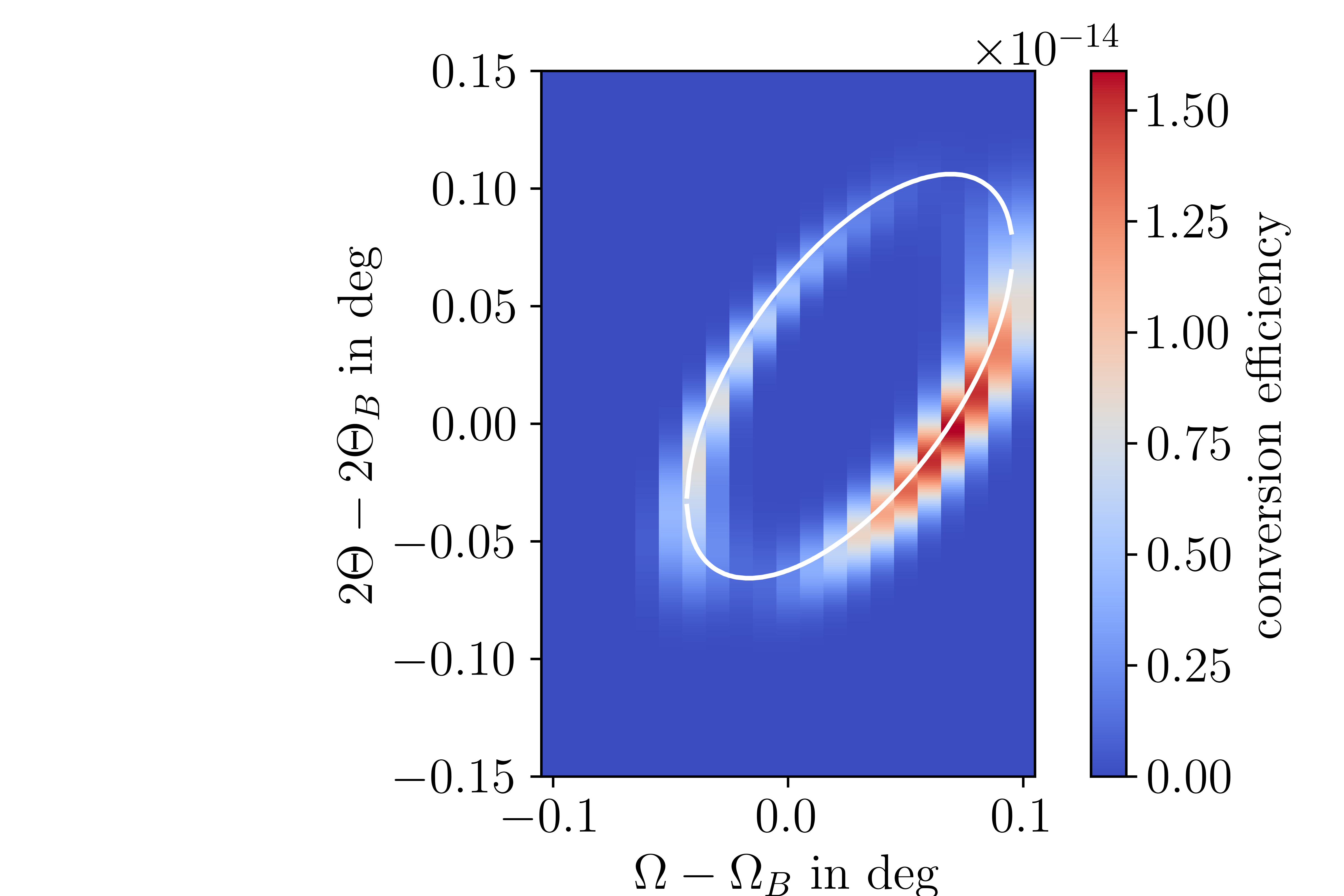}\,
   \caption{Angular space map of XPDC conversion efficiency as measurable under experimental conditions (angular acceptance: $57 \text{ mdeg} \times 1 \text{ mdeg}$---\mbox{hor. $\times$ ver.}, spectral acceptance $1 \text{ eV}$ (FWHM)). The signal exhibits a distinct elliptical signature, which results directly from momentum conservation. A simplified trace corresponding to the central pump, signal and idler energies is superimposed in white.}
   \label{fig:PM-ellipse}
   \end{center}
\end{figure}

In our next step, we aim to predict the outcome of such experiments based on our DFT-model.
To this end, we do not only require knowledge of $\mvec{K}$ though, but also of ${\ZH}^{(1)}_{\textsc{pump}}$ and $\bar{C}^{(1)}_{\textsc{vac}}$---these would complete Eq.~(\ref{eq:observable_kw}).
The optical vacuum fluctuations characterized by $\bar{C}^{(1)}_{\textsc{vac}}$ are of particular importance. They couple to the material's nonlinearity and enable the creation of optical idler photons.
As such, the role of these fluctuations can be seen in direct analogy to usual spontaneous emission, e.g., fluorescence of atoms or molecules. Just as the fluctuations enable spontaneous (linear) emission, they likewise enable the spontaneous nonlinear conversion (i.e., XPDC).
Evaluating the strength of these fluctuations would be straightforward, if the process occurred truely \emph{in vacuo}. Instead, we are of course interested in XPDC inside a sample, where the dielectric properties of the material influence electromagnetic fields. This holds for external fields, which would be re-shaped by the dielectric response within the sample, as well as for internal field fluctuations.
In fact, both aspects can be linked via the fluctuation-dissipation theorem \cite{1966Kubo-FD_theorem}---as has been shown by Landau for instance \cite{1980Landau-BOOK9.2}:
%%%
\begin{align}
  \langle
  (\Aqu[]{}{\mvec{x}^\prime,t^\prime})_\nu \, (\Aqu[]{}{\mvec{x},t})_\mu
  \rangle_{\textsc{vac}}
  &=
  \frac{-1}{\pi} \int \!\! d\omega ~ \expo[]{\omega (t^\prime - t)} ~
  \text{Im}[{D}^R_{\nu\mu}(\omega;\mvec{x}^\prime,\mvec{x})]
  .
  \label{eq:FD-Landau}
\end{align}
Using Landau's notation, ${D}^R_{\nu\mu}$ signifies the retarded electromagnetic Green's function inside the medium 
\footnote[6]{The presented treatment of field fluctuations by means of the fluctuation-dissipation theorem has seen several applications. For the case of spontaneous (linear) emission in a dielectric medium, it was used by Barnett et al. \cite{1992Barnett-SpontaneousEmissionDielectric} for example, while Klyshko established this technique for various long-wavelength regimes of parametric conversion \cite{1988Klyshko-BOOK}.}.
For our case of interest, this takes on a paricularily simple form given that diamond is practically isotropic at optical wavelength. The  full Fourier transform of Eq.~(\ref{eq:FD-Landau}) yields the desired correlation function
\begin{align}
  [\bar{C}^{(1)}_{\textsc{vac}} \argline{\prime}{i}{\prime}{i}{}{i}{}{i}]_{\nu\mu}
  &= 
  (2\pi)^5 \, \delta^3(\mvec{k}_i^\prime - \mvec{k}_i) \, \delta(\omega_i^\prime - \omega_i) \,  
  \Big( \delta_{\nu\mu} - {(\mvec{k}_i)_\nu (\mvec{k}_i)_\mu}/{\abs{\mvec{k}_i}^2} \Big) \, 
  (-4) \, \text{Im}\Big[ \frac{1}{\omega_i^2 \, \alpha^2 \, {\varepsilon}(\abs{\omega_i}) - \abs{\mvec{k}_i}^2 } \Big]
  .
  \label{eq:C_in_diel}
\end{align}
Here, ${\varepsilon}(\abs{\omega})$ denotes the dielectric function of the sample.\\
Notably, $[\bar{C}^{(1)}_{\textsc{vac}}]_{\nu\mu}$ is diagonal both in frequency and wavevector 
\footnote[7]{Being diagonal in frequency and wavevector implies that $[\bar{C}^{(1)}_{\textsc{vac}}]_{\nu\mu}$ is translationally invariant in time and space. The first invariance is a typical property of equilibrium correlation functions. The fluctuations are not bound to any specific (absolute) reference time; instead, only their relative timing is relevant. Similarily---if the medium is considered to be homogeneous---the spatial dependence is not fixed to an absolute reference point, either.}, 
which allows for further simplification of Eq.~(\ref{eq:observable_kw}).
We can write much more compactly now
\begin{align}
  \frac{d P_{\textsc{xpdc}}(\mvec{k}_s)}{d\Omega_s d\omega_s}
  &=
  {2 \,\alpha^6 \, l_w \, \omega_s (\polvec{s})_\sigma  (\polvec{s}^*)_\rho} \,
  \int \!\! d\omega_i  \!\!\! \int \!\! d^3k_i  ~  
  \frac{1}{\omega_s + \omega_i} \,
  \Phi_{\sigma\rho}(\mvec{k}_i + \mvec{k}_s - \mvec{G},\omega_s + \omega_i) ~\nonumber \\
  &\times
  \text{Im}\Big[ \frac{-4}{\omega_i^2 \, \alpha^2 \, {\varepsilon}(\abs{\omega_i}) - \abs{\mvec{k}_i}^2 } \Big] ~
  \frac{1}{V_\diamond^2} ~ \Big( \abs{\mvec{K}_{\diamond}(0, \mvec{G}, -\omega_i)}^2 - \abs{\mvec{k}_i \cdot \mvec{K}_{\diamond}(0, \mvec{G}, -\omega_i)}^2 / \abs{\mvec{k}_i}^2 \Big)
  ,
  \label{eq:observable_simpl_J}
\end{align}
where all dependence on the incident x-rays could be collected into the diagonal quantity
\begin{align}
  \Phi_{\sigma\rho}(\mvec{k},\omega) 
  &=
  \frac{1}{(2\pi)^{11}} \, \frac{\alpha \omega}{2 \, l_w}
  \int \!\! d^3k_p \!\!\! \int \!\! d^3k_p^\prime ~   
  \tilde{w}(\mvec{k} - \mvec{k}_p) \,  
  \left( \tilde{w}(\mvec{k} - \mvec{k}_p^\prime) \right)^* ~
  [{\ZH}^{(1)}_{\textsc{pump}} \argline{\prime}{p}{}{}{}{p}{}{}]_{\sigma\rho}
  \label{eq:xpdc-new-J}
  .
\end{align}
It is normalized such that for an x-ray pulse, which has both well-defined polarization and beam-like propagation behavior, it gives the distribution of its total number of photons $N_{ph}$ across its spectrum ($d\omega$) and wave-vector spread ($d^3k$). For (quasi-)continuous sources, the photon number may be exchanged for a rate. This renders $\Phi_{\sigma\rho}(\mvec{k},\omega)$ a differentially resolved flux, while the observable becomes a scattering rate, rather than a yield.
Based on this notion, we adopt a simple model for $\Phi_{\sigma\rho}(\mvec{k},\omega)$ featuring a Gaussian spectrum of width $\Omega_{p0}$ centered around the photon energy $\omega_{p0}$. Representing a well-collimated x-ray beam,  we restrict the divergence to a narrow cone that is defined by a Gaussian spread of transverse momentum components of width $1/\delta_{p0}$
\begin{align}
  \Phi_{\sigma\rho}(\mvec{k},\omega) 
  &=
  \frac{N_{ph} \, \delta_{p0}^2 \, c}{(2\pi)^{3/2} \, \Omega_{p0}} \, (\polvec{p0})_\sigma  (\polvec{p0}^*)_\rho ~
  e^{-({\omega} - \omega_{p0})^2 / 2\Omega_{p0}^2} ~
  e^{-(\mvec{k}^{\bot})^2 \delta_{p0}^2 /2} ~
  \delta(\omega - c \abs{\mvec{k}})
  \label{eq:J-model}
  .
\end{align}
Here, the Dirac-$\delta$ is used to encode the x-rays' dispersion relation, while the prefactor ensures overall normalization to $N_{ph}$.
Despite its simplicity, the model gives a good representation of synchrotron and FEL sources used to drive nonlinear conversion 
\footnote[8]{It should be noted that the model's (Gaussian) emphasis on the center of the distribution is sufficient to compute the bulk of nonlinear conversion.
However, it should not be applied to assess concurrent \emph{linear} background effects. As their conversion efficiencies are orders of magnitude larger, they will produce significant scattering even from the comparatively weak tails of the incident distribution. More realistic descriptions for linear processes should thus account for Lorentzian-type spectra (see also below XXX).}.
If necessary, more detailed descriptions of the incident field can be obtained from measured distributions \cite{2012Singer-thesis,2011Vartanyants-CohLCLS} or start-to-end simulations \cite{2013Chubar-SRWWavefront,2014Klementiev-XRT,2016Samoylova-WPG}.\\
Using Eq.~(\ref{eq:J-model}), we can evaluate our XPDC-observable.
This becomes particularily straightforward for cases, when the idler photon's energy $\omega_i$ remains below the band gap.
Without significant optical absorption, the dielectric function then becomes real $\varepsilon \rightarrow n^2$, such that the optical dispersion relation reduces to another $\delta$-function $\propto \delta(n(\omega_i) \, \omega_i - c \abs{\mvec{k}_i})$ involving the real-valued refractive index.
Performing all integrations with the ensuing constraints, we arrive at
\begin{align}
  \frac{d P_{\textsc{xpdc}}(\mvec{k}_s)}{d\Omega_s d\omega_s}
  &=
  \Big( \frac{d\sigma}{d\Omega_s} \Big)_{\text{Th}} 
  \frac{\alpha \, \sqrt{8\pi} \,N_{ph}  \,  l_w \, \omega_s}{ \Omega_{p0} \, \omega_{p0}}  \,
  e^{-(\omega_s + \frac{c}{n} \, \abs{\mvec{k}_{i0}} - \omega_{p0})^2 / 2\Omega_{p0}^2}  ~ 
  \frac{1}{\abs{{k}_{i0}^{\shortparallel} - n \abs{\mvec{k}_{i0}} }}  \nonumber \\
  &\times
  \frac{1}{V_\diamond^2} ~ \Big( \abs{\mvec{K}_{\diamond}(0, \mvec{G}, - {c}\, \abs{\mvec{k}_{i0}} /n )}^2 - \abs{\mvec{k}_{i0} \cdot \mvec{K}_{\diamond}(0, \mvec{G}, - {c} \, \abs{\mvec{k}_{i0}}/n )}^2 / \abs{\mvec{k}_{i0}}^2 \Big)
  .
  \label{eq:xpdc-final}
\end{align}
First of all, we note that our result is proportional to the Thomson cross section $({d\sigma}/{d\Omega_s})_{\text{Th}}$. This is a characteristic trait of non-resonant scattering phenomena, of which XPDC is a nonlinear variant. Beyond this, our result scales with the length of the interacting sample $l_w$ and---quintessentially---with the nonlinear response function discussed above. It's twofold structure of the form \mbox{$\abs{\mvec{K}_{\diamond}}^2 - \abs{\smash{\hat{\mvec{k}}_{i0} \cdot \mvec{K}_{\diamond}}}^2$} results from the transverse vectorial coupling to the idler photon 
\footnote[9]{As a side remark, we want to point out that the transverse coupling to the idler field should imply insensitivity to longitudinally coupling phenomena. In particular, we expect XPDC not to couple to bulk plasmons, which are inherently longitudinal excitations. This observation stands in marked contrast to recent experimental claims \cite{2019Borodin-SharonPlasmon}}. 
This transversality constraint was formally apparent in Eq.~(\ref{eq:C_in_diel}) and can be seen as an imprint on the nonlinear scattering pattern (cf. Fig.~(\ref{fig:2D-pattern})).
Finally, the refractive index was effectively fixed at the central idler energy $n = n(\omega_{p0} - \omega_s)$, while the idler photon's momentum followed from the given constraints to be
%%
%\begin{align}
%  \mvec{k}^{\bot}_{i0} 
%  &= 
%  \mvec{G}^{\bot} - \mvec{k}^{\bot}_s \\
%  %
%  {k}_{i0}^{\shortparallel} 
%  &=
%  \frac{1}{n^2 - 1} ~ \Big( n^2 \alpha A - \sqrt{\abs{\mvec{k}^{\bot}_i}^2 (n^2-1) + n^2\alpha^2 A^2} \Big)  \\
%  %
%  \text{with: }
%  A
%  &=
%  \omega_s - c ({k}_s^{\shortparallel} - {G}^{\shortparallel}) \nonumber
%  .
%\end{align}
%%
%
\begin{align}
  \mvec{k}^{\bot}_{i0} 
  &= 
  \mvec{G}^{\bot} - \mvec{k}^{\bot}_s \\
  {k}_{i0}^{\shortparallel} 
  &=
  \frac{1}{n^2 - 1} \, \Big( n^2 \alpha (\omega_s - c ({k}_s^{\shortparallel} - {G}^{\shortparallel})) - \sqrt{\abs{\mvec{k}^{\bot}_i}^2 (n^2-1) + n^2\alpha^2 (\omega_s - c ({k}_s^{\shortparallel} - {G}^{\shortparallel}))^2} \Big)  
  .
\end{align}
These relations correspond to the conservation of the average momenta. Using the terminology of nonlinear optics, this corresponds to the so called ``phase-matching condition'' for XPDC.\\
Using Eq.~(\ref{eq:xpdc-final}), we can achieve our stated goal to simulate nonlinear scattering patterns.
In doing so, we will focus on the parameter range covered within our subsequent discussion of experimental results.
We take the incident spectrum to be centered around \mbox{$\omega_{p0} = 412.2 \text{ a.u. } (\sim 11216 \text{ eV})$} with a width of $1 \text{ eV}$ (FWHM). The detection is set to an energy $\omega_s$ that is detuned downwards by  $5 \text{ eV}$.
For the time-being, we set the number of incident photons equal to unity---thus, all results can be thought of as conversion efficiencies.
The diamond sample is taken to be $500 ~\mu\text{m}$ thick and oriented for the (400) Bragg reflection at the fundamental photon energy $\omega_{p0}$. The corresponding Bragg angle is $\theta_B = 38.304 \text{ deg}$.
For different rocking angles, i.e., rotations of the sample inside the scattering plane, we obtain the nonlinear scattering patterns shown in Fig.~\ref{fig:2D-pattern}.\\
In two dimensions, we map the in-plane scattering angle $2\theta$ and out-of-plane scattering angle $\chi$ for down-converted photons of \mbox{energy $\omega_s$}. 
As scattering plane, we define the vertical plane containing the central incident wave vector ($\mvec{k}_{p0}$). The surface normal of this plane is given by the horizontal polarization vector of the undulator source.
The views are centered around the fundamental's Bragg position at $2\theta = 2\theta_B$ and $\chi = 0 \text{ deg}$.
Each pixel gives the nonlinear conversion efficiency for the solid angle that it covers\footnote[10]{Besides integrating over an element of solid angle, we also integrate over an energy interval $\Delta \omega_s$ corresponding to $\sim 0.05 \text{ eV}$.} ($1 \text{ mdeg} \times 1 \text{ mdeg}$).\\
We find that the XPDC signal is distributed in a circular pattern for most of the rocking region. This is in agreement with simple considerations based purely on momentum algebra.
In addition, our calculations predict a superimposed intensity modulation that appears as a direct consequence of the transverse coupling inherent in Eq.~(\ref{eq:xpdc-final}), viz. $\propto \abs{\mvec{K}_{\diamond}}^2 - \abs{\smash{\hat{\mvec{k}}_{i0}} \cdot \mvec{K}_{\diamond}}^2$.
For illustration of its effect, consider the extremal case when $\mvec{K}_{\diamond}$ is parallel to the unit vector $\hat{\mvec{k}}_{i0}$. Then, the whole expression vanishes exactly and XPDC is fully suppressed. More intermediate configurations account for the gradual modulation around the circle (see Figs.~\ref{fig:2D-pattern} b and c).
To our knowledge, this characteristic signature has not been pointed out previously for XPDC and could potentially be used to unambiguously identify the effect.\\
Approaching the edges of the rocking range at $\Omega - \Omega_B = \text{-$50$ and +$100$ mdeg}$ (cf. Figs.~\ref{fig:2D-pattern} a and d, respectively), the circular scattering pattern collapses into a single, point-like feature. These loci of concentration allow for particularily efficient collection of signal photons, as Kleinman \cite{1968Kleinman-TheoryOPN} and later Freund and Levine \cite{1969FreundLevine-parametricconv} point out using the term ``edge enhancement''.\\
In order to render our results more directly comparable to the experimental study of Sec.~\ref{sec:experiment}, we should further convolute the 2D patterns across the spectral and angular resolution of the employed analyzer setup.
Assuming again a passwidth of $1 \text{ eV}$ (FWHM), vertical resolution of $1 \text{ mdeg}$ ($2\theta$) and indiscriminate integration across the central $57 \text{ mdeg}$ horizontally ($\chi$), we arrive at conversion efficiencies as ``seen'' by the apparatus.
Stacking the convoluted results with respect to scattering angle $2\theta$ and rocking angle $\Omega$, we obtain a new 2D map in angular space, which we plot in Fig.~\ref{fig:PM-ellipse}.
The map exhibits a distinct elliptical signature, along which the XPDC signal appears. Once again, we can reproduce this behavior from simple momentum algebra, using
\begin{align}
  &\mvec{k}_p + \mvec{G}
  =
  \mvec{k}_s + \mvec{k}_i 
  &&\omega_p
  =
  \omega_s + \omega_i
\end{align}
as well as their respective dispersion relations.
The resulting phase-matching condition is drawn as a white line in Fig.~\ref{fig:PM-ellipse}.
We want to emphasize that this signature is a fundamental characteristic of XPDC and is therefore ideally suited to identify the effect.\\
Finally, we observe that the effect of ``edge enhancement'' is less pronounced in the present configuration---though still perceptible.
Overall, the resulting conversion efficiencies in the range of $10^{-14}$ are remarkably low, which poses significant challenges for the experimental observation of XPDC.
Claims to the contrary---as seen in some recent publications \cite{schori2017parametric,borodin2017high,sofer2019observation,2019Borodin-SharonPlasmon}---should accordingly be met with caution. 

%%%%%%%%%%%%%%%%%%%%%%%%%%%%%%%%%%%%%%%%%%%%%%%%%%%%%%%%%%%%%%%%%%%%%%%%%%%%%%%%%%%%%%%%%%%%%%%%%%%%
\section{Experiment}
\label{sec:experiment}
\begin{figure}[h!]
\begin{center}
\includegraphics[width=.5\textwidth]{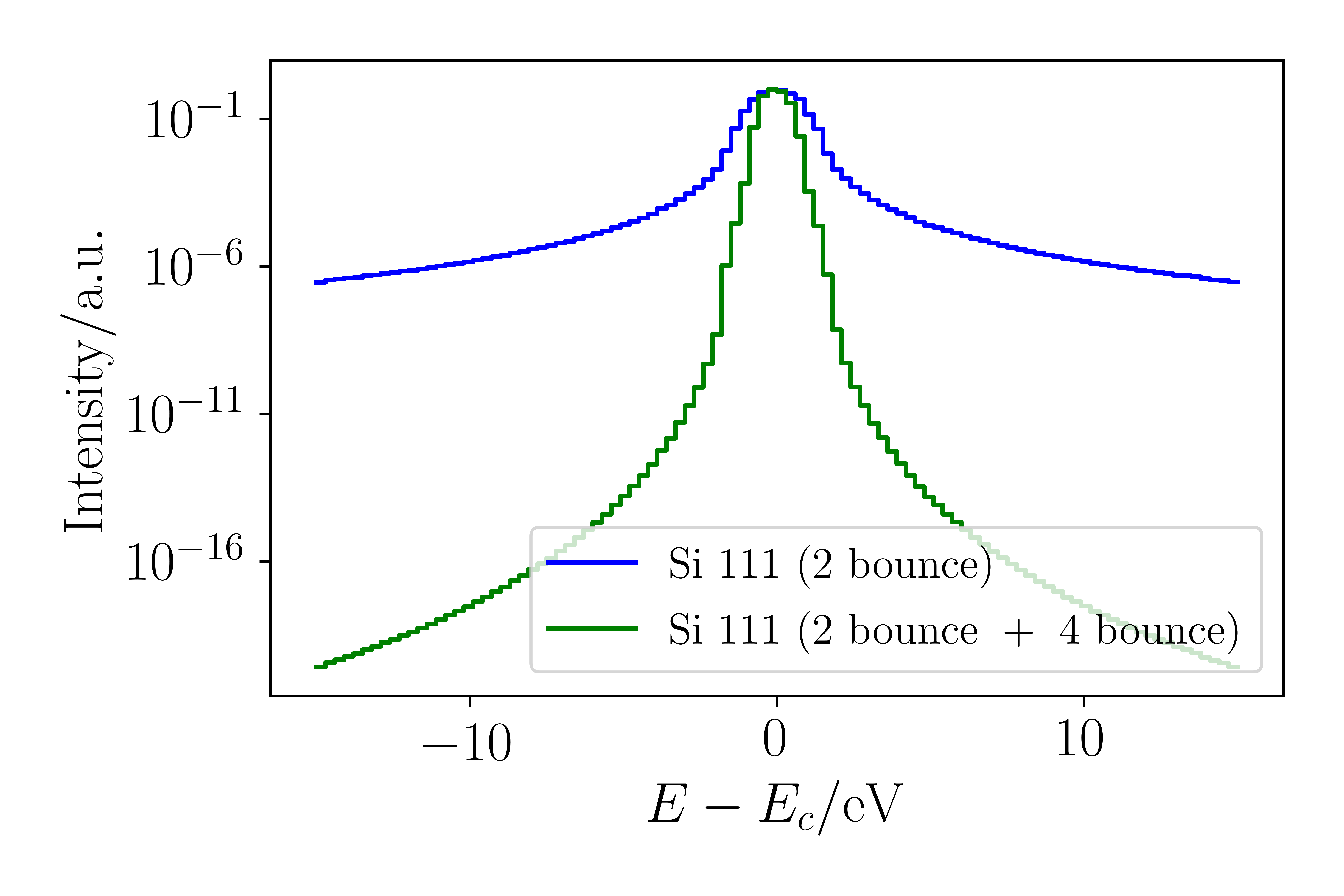}
\caption[0.5\textwidth]{OASYS simulations of the spectral intensity distribution after a conventional Si 111 two-bounce monochromator in comparison with an additional Si 111 four-bounce monochromator. The spectral tails of the incident undulator radiation are suppressed by several orders of magnitude by the additional Bartels monochromator, while the bandwidth of 1 eV remains almost constant.}
\label{fig:4-bounce}
\end{center}
\end{figure}

For the conversion of hard x-rays into visible photons the theoretical model predicts conversion rates systematically below $10^{-13}$. 
For these photon-hungry effects highly intense x-ray sources such as 3rd generation synchrotrons are indispensable.
But even with sufficiently high flux available, the resolution of the predicted scattering pattern of the nonlinear processes requires high angular precision in addition to energy discrimination.\\
Phase-matching in this regime is achieved for sample angles $\Omega$ close to the Bragg angle of the pump field $\theta_B$.
The scattering angle of the converted radiation $2\theta_s$ alike differs only by a narrow margin from the scattering angle of the elastically scattered radiation $2\theta_B$. 
In order to observe the characteristic scattering signature (see Figure \ref{fig:PM-ellipse}) and to obtain the required angular resolution, a $\Omega -2\theta$ (2-circle) diffractometer is used. 
% the following sentence should describe our envisioned measurement; we were not able to use the setup for this 
To map the sample angle specific scattering features (cf. Figure \ref{fig:2D-pattern}) a rotation of the detector arm out of the scattering plane ($\chi$) is needed.
In addition to precise angular movement, a constraint of the detected solid angle is required.\\
Furthermore, the experimental setup for investigation of XPDC requires an energy discrimination for the scattered radiation.
The energy difference between fundamental radiation and converted signal is small with regard to the pump energy. It yields the idler energy in the regime of $\hbar \omega_i= 5$ eV.
Because of this minor energy difference, the incident and scattered radiation need to be spectrally as pure as possible. \\
In general, when referring to monochromatized x-radiation at modern synchrotron sources, the usage of a conventional Si 111 double crystal monochromator (DCM) is implied. Due to the relatively broad emission spectrum ($\sim 30$ eV (FWHM)) and the finite width of Darwin curves, the transmitted spectrum by the DCM is extended and includes spectral components, which are equivalent to the converted signal. 
These spectral tails equally spread to both sides of the spectrum and behave in a Lorentzian manner, extending over a long spectral range with flanks decreasing only be $1/(E- E_c)^n$, where $n$ indicates the number of reflections.
% Include conceptual difference of XPDC in optical and x-ray regime (CHECK ME)
This special spectral behaviour of x-ray sources exposes the inherent difference between the x-ray and optical regime. The latter provides sources with distinctly narrower bandwidth, such that down-converted energies are not present in the incident pump field to a substantial degree.  \\
To be able to observe the weak effect of parametric conversion of x-rays into signal photons, which are detuned by only a few eV, the contribution of the same spectral tails stemming from the incident beam need to be reduced as much as possible.
For this reason we designed a high-spectral-suppression setup, which is based on additional crystal optics.\\
In addition to the two-bounce monochromator (DCM) of conventional setups, the undulator spectrum is filtered by a four-bounce monochromator in Bartels geometry (cf. Figure \ref{fig:p09setup}).
This geometry with four crystal reflections effectively reduces the intensity of the spectral tails by several orders of magnitude, while keeping the throughput bandwidth constant at 1 eV (FWHM). At the same time it constraints the divergence of the incident radiation.\\
Likewise, we employ a four-bounce analyzer behind the sample to discriminate the down-converted signal from the elastically scattered radiation (Figure \ref{fig:p09setup}).
The choice of crystals is motivated by the aim to achieve a strong suppression of the spectral tails on the one hand, while providing a relatively broad bandwidth throughput on the other hand. This is enabled by silicon channel-cut crystals, which have a 111 surface cut. 
This orientation yields the broadest bandwidth with $1.3$ eV (FWHM) for hard x-rays in the regime of 11 keV. 
Both, the additional monochromator and the analyzer stage, are assembled by two consecutive channel-cut crystals. In contrast to four independent crystals, the use of two channel-cut crystals reduces the degrees of freedom for optimization, but eases the general alignment procedure (which is an advantage for non-permanent setups).\\
The conventional (2-bounce) and extended (2+4-bounce) monochromator geometries were simulated by the ray tracing code OASYS \cite{rebuffi2017oasys} (Figure \ref{fig:4-bounce}) to obtain an estimate on the expected improvement.

%%%%%%%%%%%%%%%%%%%%%%%%%%%%%%%%%%%%%%%%%%%%%%%%%%%%%%%%%%%%%%%%%%%%%%%%%%%%%%%%%%%%%%%%
\subsection{Setup and Calibration}
\label{ssec:setup_cal}
\begin{figure}[h!]
\begin{center}
\includegraphics[width=0.75\textwidth]{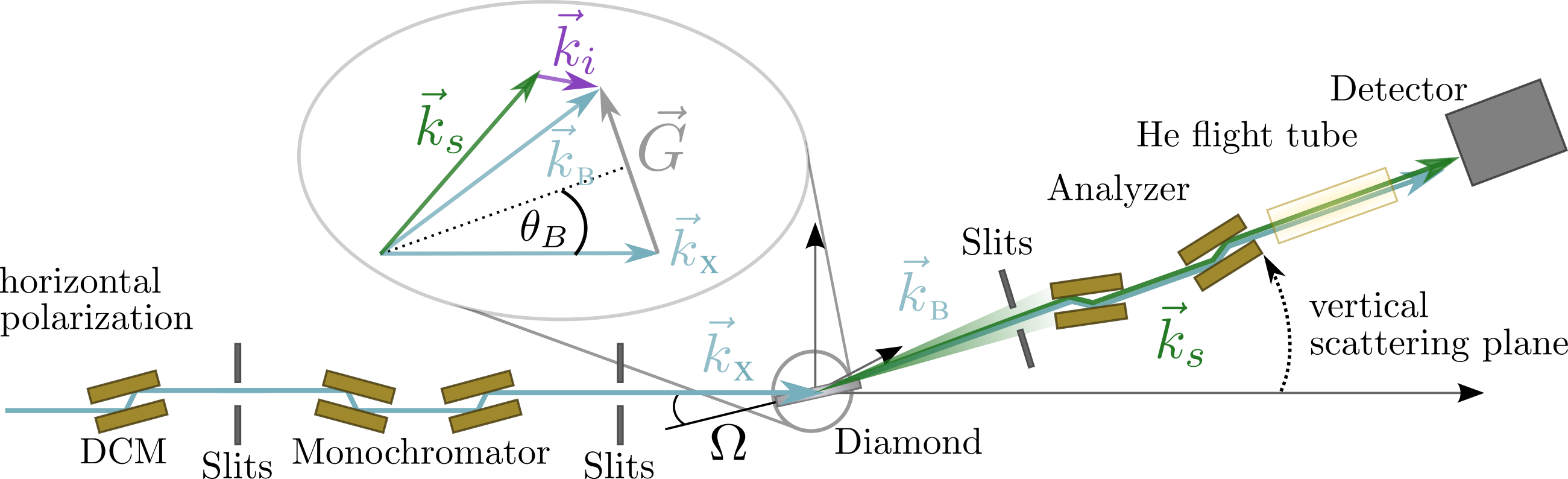}
\caption[0.5\textwidth]{Experimental setup for observing the characteristic scattering pattern of XPDC to VIS. After the beamline monochromator, an additional monochromator stage is used to further suppress elastic tails of the incident spectrum. The diamond sample crystal scatters in the vertical plane (perpendicular to the plane spanned by incident beam and polarization). The detection of the scattered photons is performed by a 4-bounce analyzer and the photon counting detector.}
\label{fig:p09setup}
\end{center}
\end{figure}

\begin{figure}[h!]
\begin{center}
\includegraphics[width=.5\textwidth]{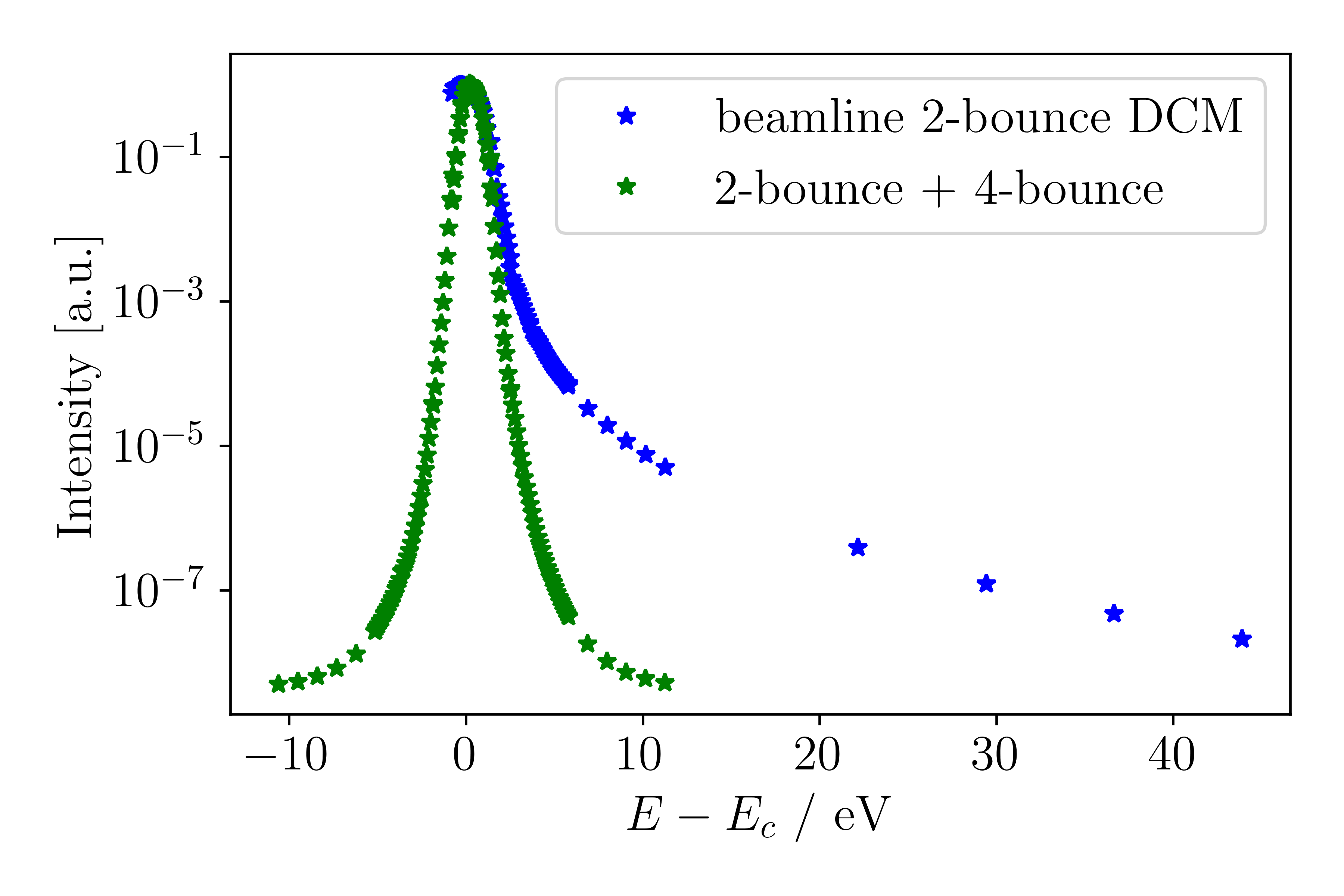}
\caption[0.5\textwidth]{Measurements of the spectral intensity distribution after a conventional Si 111 two-bounce monochromator in comparison with an additional Si 111 four-bounce monochromator. For an energy offset of $5$ eV an additional suppression by 3 orders of magnitude is achieved.}
\label{fig:4-bounce_exp}
\end{center}
\end{figure}

\begin{figure}[!tbp]
  \centering
  \includegraphics[width=0.49\textwidth]{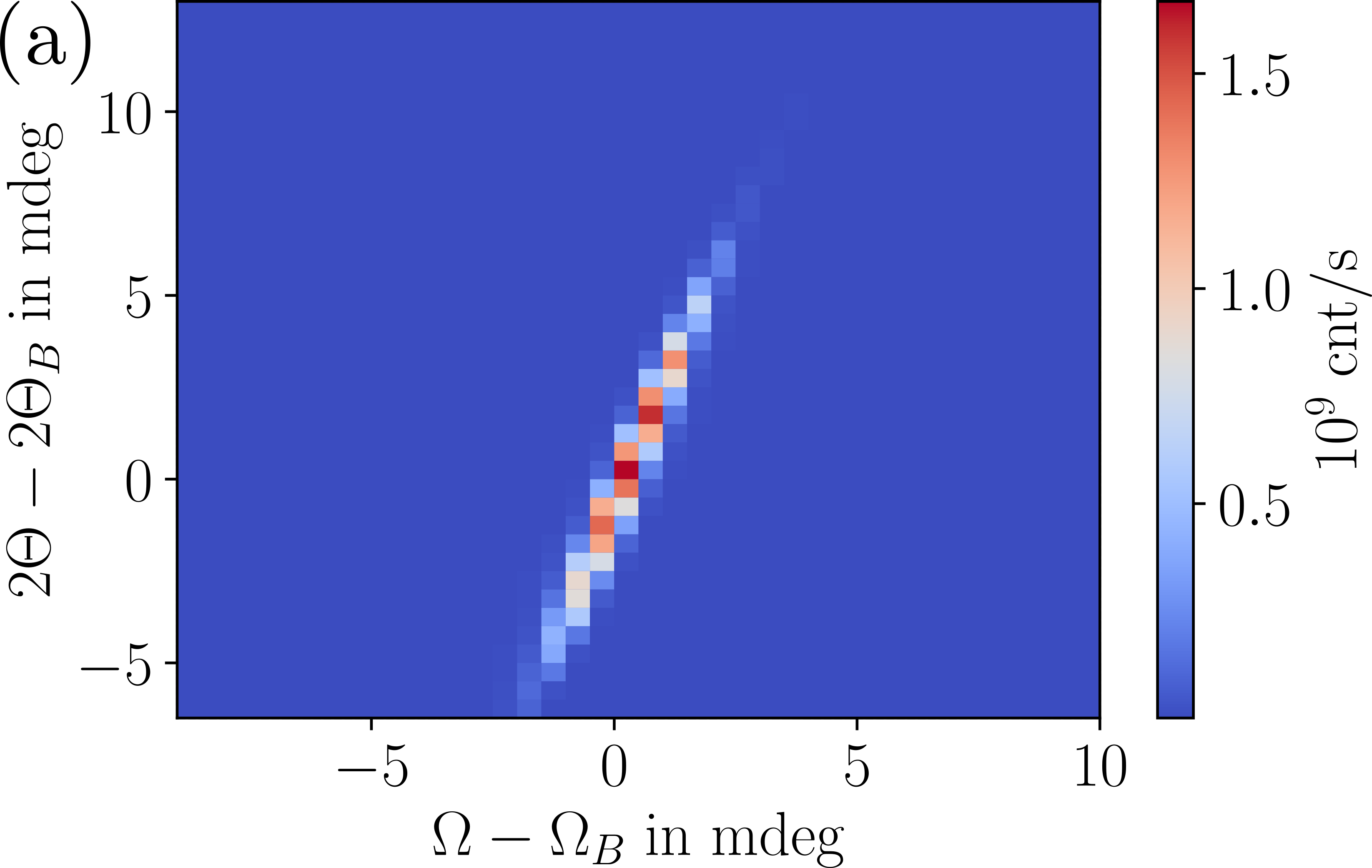}\label{fig:f1}
  \hfill
  \includegraphics[width=0.49\textwidth]{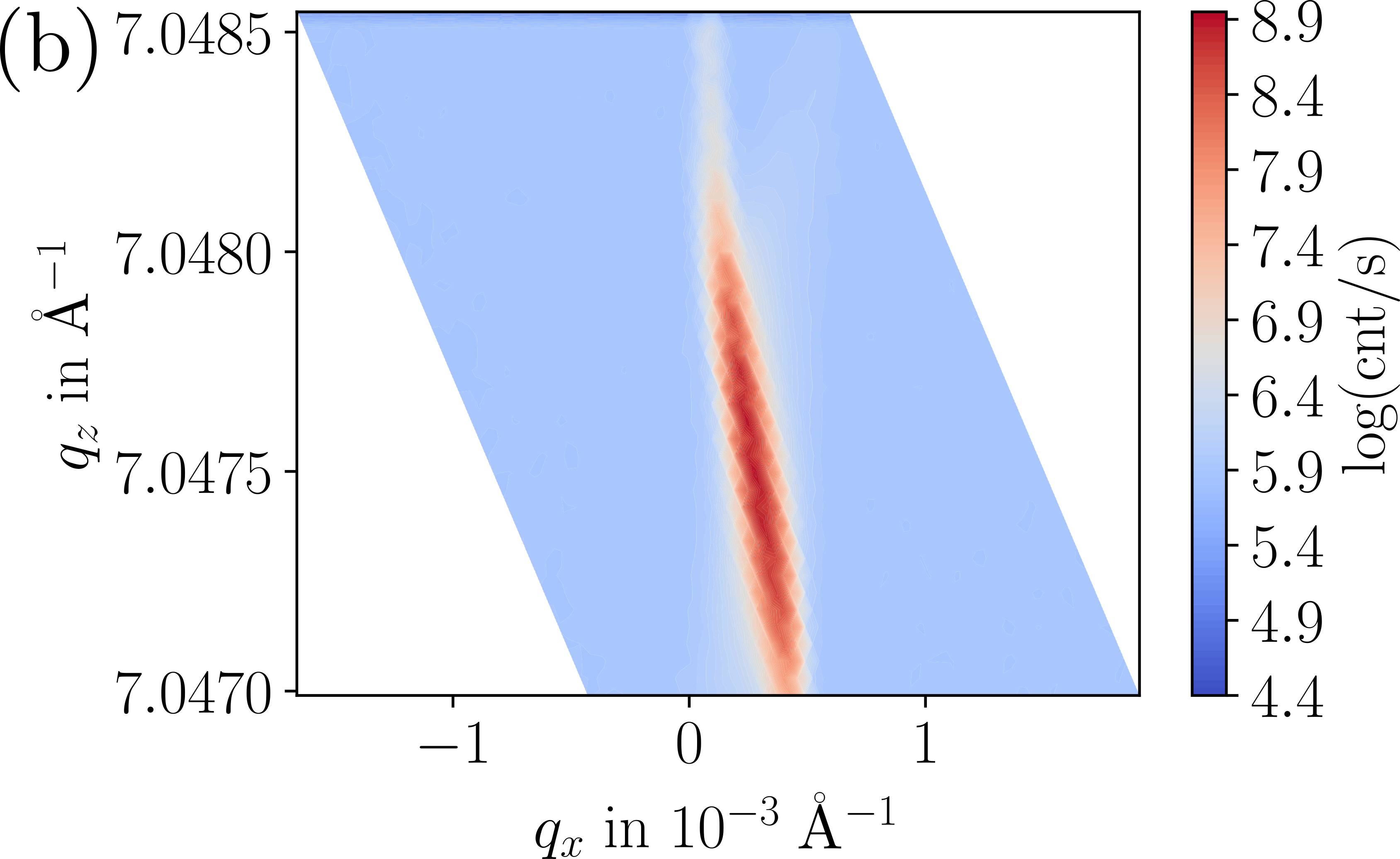}\label{fig:f2}
  \caption{Instrumental function in angular and reciprocal coordinates. The sample angle $\Omega$ for 400 orientation and the detector angle $2\theta$ are scanned in a range of $\pm 10$ mdeg. The scattering features of the instrumental function have small extension in angular and reciprocal space.}
  \label{fig:RSM_undetuned}
\end{figure}

A synchrotron beamline, which provides sufficient flux in combination with a diffractometer providing the required degrees of freedom is the P09 beamline at Petra III, at which the experimental campaign was conducted. 
The complete experimental setup is shown in Figure \ref{fig:p09setup}: The 11.216 keV beam is monochromatized by the DCM and the additional monochromator stage. It impinges onto the sample at an angle $\Omega$. The diamond crystal, which is chosen for its high crystallographic quality,acts as the nonlinear medium for down-conversion. A $500\ \mu$m thick substrate is used with a 100-surface cut in vertical reflection geometry.  
The scattered radiation is analyzed by a four-bounce crystal configuration, similar to the additional monochromator stage. The diffracted photons are acquired with a pixel detector (lamdbda detector, $55\ \mu$m pixel size \cite{pennicard2013lambda}). 
% apertures
Apertures are located in front and after the additional monochromator and before the analyzer stage.\\
\\
% alignment procedure
During the alignment procedure, each crystal was set to fulfill the Bragg condition of the fundamental radiation and subsequently rocked to obtain the individual rocking curves. Thereby a width (FWHM) of $1.9 \pm 0.03$ mdeg for the Si 111 reflexes of the additional monochromator and analyzer crystals are measured. 
These values are close to the theoretically Darwin widths for a double-crystal of silicon 111 $\omega^{Si_{111}} = 31.818 \ \mu$rad ($1.823$ mdeg)\cite{stepanov2004x}.
These results underline the accuracy, which was achieved with the presented configuration.\\
%experimental results for mono spectra
Before using the diffraction setup to investigate the nonlinear processes the whole apparatus is carefully characterized. First, the spectral intensity distribution of the DCM and the additional four-bounce monochromator is characterized\footnote[11]{The spectrum of the DCM is measured by the successive four-bounce monochromator, while the spectrum of the additional monochromator is measured by the analyzer configuration.} (Figure \ref{fig:4-bounce_exp}).
The simulated bandwidth of 1 eV (FWHM) is reproduced by the measurement for the DCM and the additional monochromator stage.
Yet, the suppression of the spectral tails was only improved by three orders of magnitude for an energy offset of $5$ eV. 
This discrepancy between simulation and experiment is mainly caused by the neglect of experimental uncertainties in the underlying simplified simulation. Thermal effects, for example, are completely ignored, as well as diffuse scattering from the crystal surfaces, which can not entirely be avoided.
% Positive statement
Nevertheless, the improvements over the standalone DCM are significant and provide improved conditions for observing XPDC.
\\
% Instrumental Function
In a next step, the instrumental function of the complete experimental setup is measured. For that purpose, the DCM, the additional monochromator and analyzer stage \footnote[12]{The monochromator and analyzer are calibrated, such that a direct correlation between rotation angle and selected energy is available.} are optimized for the fundamental energy of 11.216 keV and the sample angle of the diamond crystal\footnote[13]{The rocking curve of the diamond in 400 orientation yielded $0.5$ mdeg. For a single reflection of diamond 400 the Darwin width is $\omega^{C_{400}}(11\ \text{keV}) = 1.2363\ \mu$rad ($0.071$ mdeg)} is set to the corresponding Bragg angle $\theta_B^{C_{400}}$.\\
Detailed information on the instrumental function is obtained by $\Omega - 2\theta$-scans, i.e., scans of the sample and detector angle in a narrow regime around Bragg condition. The DCM, the monochromator and analyzer crystals remain unchanged at their optimized position.  
The detection setup covers a solid angle of $57$ mdeg in horizontal and $1$ mdeg in vertical dimension around the scattering vector. These dimensions are determined by the aperture (3rd slit) and the angular acceptance of the analyzer crystals.\\
The results of the $\Omega - 2\theta$ scan are shown in Figure \ref{fig:RSM_undetuned}a for angular coordinates and in Figure \ref{fig:RSM_undetuned}b for reciprocal coordinates.
The latter representation is used to facilitate the identification of features, which originate from the optical elements of the setup and are well known in high-resolution diffractometry \cite{mikhalychev2015ab, neumann1994resolution, rutt1995resolution}.
\\
Only a low extension of instrumental function features in angular and reciprocal space is observed. The remaining feature is a combination of different effects. In particular, the wavelength and analyzer streak contribute to the remaining scattering pattern.  
The expansion of the remaining streaks is limited to $\pm 5$ mdeg along $\Omega$ and to $\pm 10$ mdeg along $2\theta$.
The characterization (Figure \ref{fig:RSM_undetuned}a and b) clearly shows that the aimed for improvement of the instrumental function is achieved.\\
Previous experiments \cite{boemer2020x, DissBoemer}, conducted with only with a two-bounce monochromator (Si 111) and analyzer (Si 440), yielded additional features, which extended beyond $\pm 100$ mdeg along $\Omega$ and $\pm 50$ mdeg along the scattering dimension $2\theta$.
The setup was dominated by monochromator and analyzer streaks. The latter being oriented parallel to the $2\theta$-axis, whereas the monochromator streak can be identified by the angle it sets with the analyzer streak, namely the Bragg angle of the sample $\theta_B$.\\
% Transmission function
Notable is also the high transmission the refined experimental setup provides. With an initial flux of $4\cdot 10^{12}$ ph/s \footnote[14]{after monochromatization by DCM and additional 4-bounce monochromator} and an integrated detected count rate of $1.6\cdot 10^{10}$ (Figure \ref{fig:f1}) corrected by the attenuation factor $7\cdot 10^{3}$, 
the overall attenuation after the sample and analyzer reflections amounts only to a factor of 3.
The setup is further optimized with regard to shielding. Beamline components are carefully shielded to reduce scattering contributions on the detector. For the reduction of air scattering an additional helium filled flight tube was introduced between analyzer and detector. 

%%%%%%%%%%%%%%%%%%%%%%%%%%%%%%%%%%%%%%%%%%%%%%%%%%%%%%%%%%%%%%%%%%%%%%%%%%%%%%%%%%%%%%%%
\subsection{Experimental results and discussion}
\label{ssec:exp_results}
With the calibrated and characterized setup we start the search for the nonlinear parametric conversion process into idler photons of energy $\hbar \omega_i = 5$ eV.
This energy is selected, because it is below the bandgap energy of diamond at $E_{gap} = 5.4$ eV. Above this energy, absorption sets in and the characteristic elliptical signature becomes blurred, making it harder to detect. 
Given this constraint, the energy offset of $5$ eV provides the optimum suppression of spectral tails (Figure \ref{fig:4-bounce_exp}).
% scanning procedure: mapping the angular parameter space for XPDC PM
For a systematic measurement for the predicted characteristic scattering signature (Figure \ref{fig:PM-ellipse}) the angular space is mapped out. 
\begin{figure}[h!]
\begin{center}
\includegraphics[width=.5\textwidth]{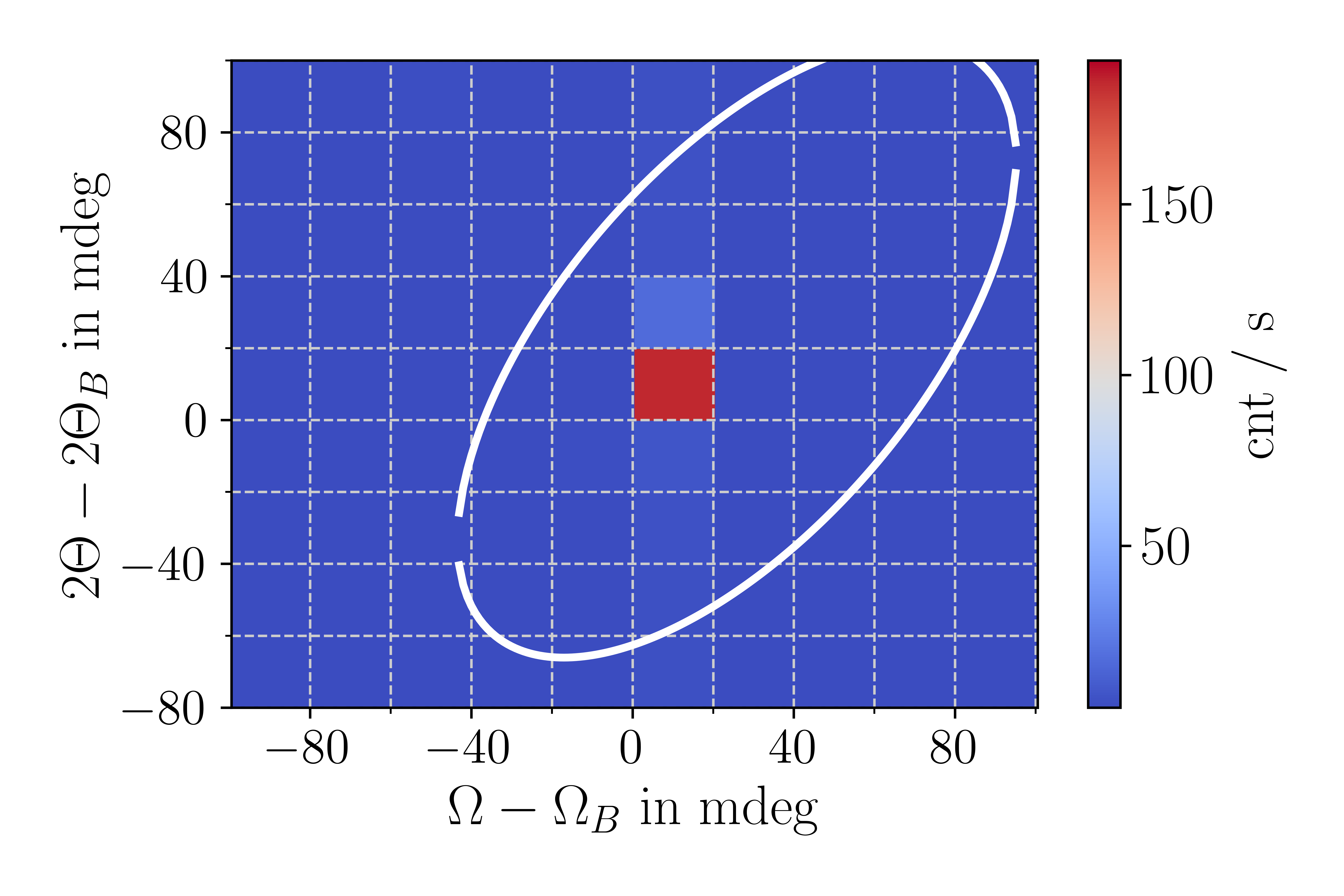}
\caption[0.5\textwidth]{Angular space map ($\Omega$ vs $2\theta$) in the regime, where phase-matching is expected for a pump energy $E_p = 11.216$ keV and idler photons of $5$ eV. The characteristic scattering signature of nonlinear parametric down-conversion is not observed.}
\label{fig:RSM_PM_5eV}
\end{center}
\end{figure}
Therefore, the sample ($\Omega$) and detector angles ($2\theta$) are scanned systematically, in a range of $\pm 100$ mdeg around the nominal Bragg angle $\theta_B$, with steps of 20 mdeg for each dimension. 
Figure \ref{fig:RSM_PM_5eV} shows the resulting angular space map, including the predicted characteristic scattering signature (white) of the nonlinear effect according to Figure \ref{fig:PM-ellipse}. 
Each tile, presents the integration of count rates\footnote[15]{For a selected region of interest on the detector. A flatfield correction is applied to the images in addition to corrections accounting for fluctuations of the incident flux.} within the covered solid angle of the detection setup, whereby its lower left corner indicates the corresponding sample and detector positions.  
The acquisition time for each position was 100 s with an incident flux at sample position of $4 \cdot 10^{12}$ ph/s \footnote[16]{This flux was measured after the DCM and additional 4-bounce monochromator directly at the sample position with a calibrated diode.}.\\
The angular space map for down-conversion of hard x-rays into $5$ eV idler photons (Figure \ref{fig:RSM_PM_5eV}) does not exhibit evidence of the expected characteristic elliptical scattering signature (white line), despite the extended integration time of 100 s accounting for the expected low conversion rate.\\
Only the central spot at Bragg condition ($\Omega = \theta_B$) shows an increased count rate of 150 cnts/s. This signal is the remainder of elastic scattering, which originates from the spectral tails transmitted by monochromators, sample and analyzer (i.e., the instrumental function).
In contrast to the elastic contributions, which were determined earlier (Figure \ref{fig:RSM_undetuned}a) for the undetuned analyzer, this elastic remainder is much weaker, since the analyzer is detuned by $5$ eV from the fundamental (pump) wavelength.\\
For sample and detector angles detuned from Bragg condition, the integrated count rate quickly decreases to the background level of below $50$ cnt/s. 
Especially for angular configurations, where phase-matching (white line) is expected, no increase in the count rate is observed. In particular, no elevated signal is recorded for positions, where the signal should be strongest according to theoretical predictions (Figure \ref{fig:PM-ellipse}), for example at $\Delta \Omega = 60$ mdeg and $\Delta 2\theta = 0$ mdeg.\\
% Einordnung der Ergebnisse 
At this point we are able to directly compare the experimental results with the theoretically predicted count rates. 
Multiplying the values given in Figure \ref{fig:PM-ellipse} by the available flux of $4\cdot 10^{12}$ per second, we determine an expected count rate of up to 0.16 ph/s.
Such a low signal strength is not resolvable with the presented experimental setup, because of the background level of $50$ ph/s.\\
% Estimation new upper bound 
Despite the absence of the characteristic scattering signature, we are able to give an upper bound for the effect's conversion efficiency based on the resolution of the setup.
With the presented experimental configuration we would be able to distinguish a count rate of $50$ cnts/s above background (for the given acquisition time of 100 s). 
Considering the available flux of $4\cdot 10^{12}$ ph/s at the sample position, a signal count rate of 50 ph/s (potential minor losses along beam path neglected) yields a conversion efficiency of $1.25 \cdot 10^{-11}$, which is the determined upper bound. \\
% is it valid to neglect QE of detector, air absorption, transmission coefficients < 1 ???
This experimentally deduced upper bound is thereby directly in agreement with the theoretically predicted conversion efficiency, which is with $10^{-14}$ even three orders of magnitude lower.\\
The final proof for x-ray parametric down-conversion into visible photons remains yet outstanding. It requires the detection of the effect's characteristic scattering signature as well as the experimental determination of the conversion efficiency. 
An additional goal would be the imaging of the characteristic scattering signature that shows prominent, specific features for a set sample angle (Figure \ref{fig:2D-pattern}).
% CHECK my wording
From our experiments, which are in accordance with our presented theoretical framework, we deduce, that further investigation of the effect at synchrotron sources seems not promising due to the following reasons.
With the available (fixed) fluxes only the acquisition times could be increased to achieve higher signal count rates.  
However, the integration time would need to be increased by a factor of 300 in order to produce detectable signal count rates; yielding acquisition times of 8.3 hours. 
Second, the reduction of measured background radiation becomes increasingly difficult, such that despite increased acquisition times the signal cannot be distinguished.
Ultimately, the fundamental difficulty remains, that for the given effect no background free measurements technique is available (to date). 

\section{Controversy}
\label{sec:controversy}
The experimental observation of x-ray parametric conversion into visible photons with a diamond sample has been claimed in a previous publication \cite{schori2017parametric}. 
Moreover, a similar study reports the observation of x-ray conversion into UV photons by a standard, lab-based source \cite{borodin2017high}.
Furthermore, several related publications in high-impact journals have successively suggested and shown 'implementations' of applications \cite{ 2019Borodin-SharonPlasmon,sofer2019observation} already. 
However, we find that the underlying claim of these studies does not withstand scientific scrutiny. In our opinion, the conversion of hard x-ray photons into visible photons was not observed so far.\\
As solid unequivocal evidence for XPDC, its characteristic scattering signature needs to be observed. As this signature follows directly from the effect’s phase-matching condition (i.e., fundamental momentum conservation) and is thus expected to be present whenever XPDC is measurable.
Moreover, we demonstrated in detail, that concepts form the optical regime (where PDC is frequently applied) cannot straightforwardly be transferred to the x-ray regime. Due to the extended spectrum of the x-ray illumination function - and its successive transmittance through the setup - a background free measurement of the effect is not possible. The differentiation between signal and background can only be established on the basis of angular scans, which map out the characteristic scattering signature.\\
However, this signature has not been provided in the aforementioned references \cite{schori2017parametric, borodin2017high, 2019Borodin-SharonPlasmon, sofer2019observation}. 
Furthermore, the conversion rates reported\cite{schori2017parametric, sofer2019observation} for XPDC range between $8 \cdot 10^{-10}$  of up to $\sim 5 \cdot 10^{-5}$.
Especially the latest reports\cite{sofer2019observation} for XPDC in LiNbO3 mention conversion rates, which deviate by 6 orders of magnitude from our here presented results. 
A detailed revision\cite{boemer2020x} of the proof-of-principle study\cite{schori2017parametric} allowed for an identification of the alleged signal as elastic scattering contribution from the experimental apparatus itself (i.e., the instrumental function).\\
\\
The result of our study, namely the absence of the characteristic scattering signature thereby leads us to reject the claim of the previous report\cite{schori2017parametric} on XPDC in diamond and challenges several other publications that have claimed observation and application of XPDC in other samples\cite{borodin2017high, 2019Borodin-SharonPlasmon,sofer2019observation}.
In addition, our experimental results are corroborated by ab-initio calculations based on non-relativistic Quantum-Electrodynamics, which are in perfect agreement with the experimentally obtained upper bound of the effect's conversion efficiency.

%Main reasons to reject the claims of reported XPDC: 
%first: no scattering signature
%second: no well characterized exp apparatus (only 2-bounce monochromator, sometimes only single-bounce analyzer (lacking suppression).
%Finally: too high conversion rates 

%Its absence – on the contrary – leads us to reject the claim of Ref. [1]. Instead of XPDC, we find all features of the scattering pattern to behave like regular elastic scattering throughout our systematic investigation of the effect’s parameter space. Notably, this includes the conditions of the alleged proof-of-principle experiment. Here, our measurements can locally reproduce the signal previously interpreted as XPDC [1], while our view of the global scattering pattern clearly refutes that interpretation. These findings are judged as scientifically valid by the referees. Furthermore, our experimental results are corroborated by ab-initio calculations based on non-relativistic Quantum-Electrodynamics (publication in preparation).

\section{Summary and Outlook}
\label{sec:outlook}
% or Conclusion and Outlook --> what is preferred?
%
%- efforts improving experimental setup
%- achieved stronger suppression of elastic tails, 
%- included longer integration time (two-orders of magnitude more, than previous studies)
%- however not sufficient to observe the low-conversion effect of XPDC
Striving to provide a clear identification of parametric down-conversion of x-rays into visible photons we aim to detect the effect’s characteristic scattering signature. 
For this purpose, we have refined our experimental methodology towards a high-spectral-suppression diffraction setup, with which we systematically map the angular space. The spectral components of the incident spectrum are more strongly suppressed and the instrumental function is substantially improved. 
Yet, the resolution of the experimental apparatus was not sufficient to resolve the characteristic signature, despite of up to 3 orders better tail suppression and 2 orders of magnitude longer integration times. 
Ultimately the weak effect could not be resolved from the remaining background signal. Yet, we find the experimentally determined upper bound of the conversion efficiency to be in full agreement with rates predicted by a QED-based ab-initio framework, presented in this work.\\
\\
Despite our null result on parametric down-conversion, the pursuit of x-ray optical wave mixing techniques remains promising. In particular, much higher count rates can be expected, if conversion processes are externally driven, i.e., stimulated. For our case of XPDC, the intrinsically spontaneous conversion into idler photons relied on the presence of weak vacuum fluctuations in the respective photonic modes. Yet, if these were already populated by externally provided photons, this would stimulate the conversion significantly. While this stimulated effect---which amounts to difference-frequency generation (DFG)---has not been demonstrated yet, its corresponding up-conversion variant---namely x-ray optical sum-frequency generation (SFG)---has already been shown to be experimentally feasible\cite{glover2012x}. Moreover, recent theoretical results by Popova et al. also indicate the visibility of conversion-sidebands in wave mixing-type diffraction\cite{popova2018theory}.\\
Notably, such externally driven effects require an additional field, which implies increased experimental complexity. At the same time, this yields potential advantages in terms of extended control, e.g., the nonlinear signals can be manipulated through the optical field’s polarization, intensity, or temporal delay. We envision to employ the methodology of this work to study such dependencies---both from a theoretical and experimental point of view. Ultimately, we are optimistic to apply x-ray optical wave mixing (SFG and DFG) as a probe of valence dynamics in future experimental campaigns.

%\section{Conclusions}
%The conclusions section should come in this section at the end of the article, before the Conflicts of interest statement.

\section*{Acknowledgements}
The authors would like to acknowledge the provision of beamtime for proposal (I-20190876) at P09 at the PETRA III synchrotron and the excellent technical support of the beamline personnel at P09 by Sonia Francoual and Julian Bergtholdt.\\
Furthermore, this experimental campaign would not have been possible without the support of Berit Marx-Glowna, Horst Schulte-Schrepping and Manfred Spiwek, who provided and cut silicon crystals of highest quality for the monochromator and analyzer stages. \\
In addition, we are thankful to Yuri Shvyd'ko for lending us a precharacterized diamond sample of monochromator quality.\\
The authors thank Mika Rassola, Kallio Antti-Jussi for their support and Alke Meents and Edgar Weckert for fruitful discussion of the high-suppression diffraction setup. \\
E. Rossi is funded by the Cluster of Excellence „CUI: Advanced Imaging of Matter“ of the Deutsche Forschungsgemeinschaft (DFG) – EXC 2056 – project ID 390715994.”

%%%END OF MAIN TEXT%%%


%\balance


%%%REFERENCES%%%
\bibliography{biblio}

\providecommand*{\mcitethebibliography}{\thebibliography}
\csname @ifundefined\endcsname{endmcitethebibliography}
{\let\endmcitethebibliography\endthebibliography}{}
\begin{mcitethebibliography}{39}
\providecommand*{\natexlab}[1]{#1}
\providecommand*{\mciteSetBstSublistMode}[1]{}
\providecommand*{\mciteSetBstMaxWidthForm}[2]{}
\providecommand*{\mciteBstWouldAddEndPuncttrue}
  {\def\EndOfBibitem{\unskip.}}
\providecommand*{\mciteBstWouldAddEndPunctfalse}
  {\let\EndOfBibitem\relax}
\providecommand*{\mciteSetBstMidEndSepPunct}[3]{}
\providecommand*{\mciteSetBstSublistLabelBeginEnd}[3]{}
\providecommand*{\EndOfBibitem}{}
\mciteSetBstSublistMode{f}
\mciteSetBstMaxWidthForm{subitem}
{(\emph{\alph{mcitesubitemcount}})}
\mciteSetBstSublistLabelBeginEnd{\mcitemaxwidthsubitemform\space}
{\relax}{\relax}

\bibitem[Shen(2002)]{2002Shen_BOOK-Nonlinear_Optics}
Y.~R. Shen, \emph{The Principles of Nonlinear Optics}, Wiley-Interscience,
  2002\relax
\mciteBstWouldAddEndPuncttrue
\mciteSetBstMidEndSepPunct{\mcitedefaultmidpunct}
{\mcitedefaultendpunct}{\mcitedefaultseppunct}\relax
\EndOfBibitem
\bibitem[Mukamel(1999)]{1999Mukamel-BOOK}
S.~Mukamel, \emph{Principles of Nonlinear Optical Spectroscopy}, Oxford
  University Press, 1999\relax
\mciteBstWouldAddEndPuncttrue
\mciteSetBstMidEndSepPunct{\mcitedefaultmidpunct}
{\mcitedefaultendpunct}{\mcitedefaultseppunct}\relax
\EndOfBibitem
\bibitem[Dixit \emph{et~al.}(2012)Dixit, Vendrell, and Santra]{2012Dixit-PNAS}
G.~Dixit, O.~Vendrell and R.~Santra, \emph{Proceedings of the National Academy
  of Sciences}, 2012, \textbf{109}, 11636--11640\relax
\mciteBstWouldAddEndPuncttrue
\mciteSetBstMidEndSepPunct{\mcitedefaultmidpunct}
{\mcitedefaultendpunct}{\mcitedefaultseppunct}\relax
\EndOfBibitem
\bibitem[Kowalewski \emph{et~al.}(2017)Kowalewski, Bennett, and
  Mukamel]{2017Kowaleski-nonadiabatic_XRD}
M.~Kowalewski, K.~Bennett and S.~Mukamel, \emph{Structural Dynamics}, 2017,
  \textbf{4}, 054101\relax
\mciteBstWouldAddEndPuncttrue
\mciteSetBstMidEndSepPunct{\mcitedefaultmidpunct}
{\mcitedefaultendpunct}{\mcitedefaultseppunct}\relax
\EndOfBibitem
\bibitem[Simmermacher \emph{et~al.}(2019)Simmermacher, Henriksen, M\o{}ller,
  Moreno~Carrascosa, and Kirrander]{2019Simmermacher-coherence}
M.~Simmermacher, N.~E. Henriksen, K.~B. M\o{}ller, A.~Moreno~Carrascosa and
  A.~Kirrander, \emph{Phys. Rev. Lett.}, 2019, \textbf{122}, 073003\relax
\mciteBstWouldAddEndPuncttrue
\mciteSetBstMidEndSepPunct{\mcitedefaultmidpunct}
{\mcitedefaultendpunct}{\mcitedefaultseppunct}\relax
\EndOfBibitem
\bibitem[Young \emph{et~al.}(2018)Young, Ueda, Gühr, Bucksbaum, Simon,
  Mukamel, Rohringer, Prince, Masciovecchio, Meyer, Rudenko, Rolles, Bostedt,
  Fuchs, Reis, Santra, Kapteyn, Murnane, Ibrahim, L{\'{e}}gar{\'{e}}, Vrakking,
  Isinger, Kroon, Gisselbrecht, L'Huillier, Wörner, and
  Leone]{2018Young-XrayRoadmap}
L.~Young, K.~Ueda, M.~Gühr, P.~H. Bucksbaum, M.~Simon, S.~Mukamel,
  N.~Rohringer, K.~C. Prince, C.~Masciovecchio, M.~Meyer, A.~Rudenko,
  D.~Rolles, C.~Bostedt, M.~Fuchs, D.~A. Reis, R.~Santra, H.~Kapteyn,
  M.~Murnane, H.~Ibrahim, F.~L{\'{e}}gar{\'{e}}, M.~Vrakking, M.~Isinger,
  D.~Kroon, M.~Gisselbrecht, A.~L'Huillier, H.~J. Wörner and S.~R. Leone,
  \emph{Journal of Physics B: Atomic, Molecular and Optical Physics}, 2018,
  \textbf{51}, 032003\relax
\mciteBstWouldAddEndPuncttrue
\mciteSetBstMidEndSepPunct{\mcitedefaultmidpunct}
{\mcitedefaultendpunct}{\mcitedefaultseppunct}\relax
\EndOfBibitem
\bibitem[Schori \emph{et~al.}(2017)Schori, B{\"o}mer, Borodin, Collins,
  Detlefs, Sala, Yudovich, and Shwartz]{schori2017parametric}
A.~Schori, C.~B{\"o}mer, D.~Borodin, S.~Collins, B.~Detlefs, M.~M. Sala,
  S.~Yudovich and S.~Shwartz, \emph{Physical review letters}, 2017,
  \textbf{119}, 253902\relax
\mciteBstWouldAddEndPuncttrue
\mciteSetBstMidEndSepPunct{\mcitedefaultmidpunct}
{\mcitedefaultendpunct}{\mcitedefaultseppunct}\relax
\EndOfBibitem
\bibitem[Borodin \emph{et~al.}(2017)Borodin, Levy, and
  Shwartz]{borodin2017high}
D.~Borodin, S.~Levy and S.~Shwartz, \emph{Applied Physics Letters}, 2017,
  \textbf{110}, 131101\relax
\mciteBstWouldAddEndPuncttrue
\mciteSetBstMidEndSepPunct{\mcitedefaultmidpunct}
{\mcitedefaultendpunct}{\mcitedefaultseppunct}\relax
\EndOfBibitem
\bibitem[Sofer \emph{et~al.}(2019)Sofer, Sefi, Strizhevsky, Aknin, Collins,
  Nisbet, Detlefs, Sahle, and Shwartz]{sofer2019observation}
S.~Sofer, O.~Sefi, E.~Strizhevsky, H.~Aknin, S.~Collins, G.~Nisbet, B.~Detlefs,
  C.~J. Sahle and S.~Shwartz, \emph{Nature communications}, 2019, \textbf{10},
  1--8\relax
\mciteBstWouldAddEndPuncttrue
\mciteSetBstMidEndSepPunct{\mcitedefaultmidpunct}
{\mcitedefaultendpunct}{\mcitedefaultseppunct}\relax
\EndOfBibitem
\bibitem[Borodin \emph{et~al.}(2019)Borodin, Schori, Rueff, Ablett, and
  Shwartz]{2019Borodin-SharonPlasmon}
D.~Borodin, A.~Schori, J.-P. Rueff, J.~M. Ablett and S.~Shwartz, \emph{Phys.
  Rev. Lett.}, 2019, \textbf{122}, 023902\relax
\mciteBstWouldAddEndPuncttrue
\mciteSetBstMidEndSepPunct{\mcitedefaultmidpunct}
{\mcitedefaultendpunct}{\mcitedefaultseppunct}\relax
\EndOfBibitem
\bibitem[Boemer \emph{et~al.}(2020)Boemer, Krebs, Diez, Rohringer, Galler, and
  Bressler]{boemer2020x}
C.~Boemer, D.~Krebs, M.~Diez, N.~Rohringer, A.~Galler and C.~Bressler,
  \emph{arXiv preprint arXiv:2002.12822}, 2020\relax
\mciteBstWouldAddEndPuncttrue
\mciteSetBstMidEndSepPunct{\mcitedefaultmidpunct}
{\mcitedefaultendpunct}{\mcitedefaultseppunct}\relax
\EndOfBibitem
\bibitem[Krebs and Rohringer()]{202XKrebs}
D.~Krebs and N.~Rohringer, manuscript in preparation\relax
\mciteBstWouldAddEndPuncttrue
\mciteSetBstMidEndSepPunct{\mcitedefaultmidpunct}
{\mcitedefaultendpunct}{\mcitedefaultseppunct}\relax
\EndOfBibitem
\bibitem[Craig and Thirunamachandran(1998)]{1998CraigThirunamachandran-BOOK}
D.~P. Craig and T.~Thirunamachandran, \emph{Molecular Quantum Electrodynamics},
  Dover, 1998\relax
\mciteBstWouldAddEndPuncttrue
\mciteSetBstMidEndSepPunct{\mcitedefaultmidpunct}
{\mcitedefaultendpunct}{\mcitedefaultseppunct}\relax
\EndOfBibitem
\bibitem[Oddershede \emph{et~al.}(1984)Oddershede, Jørgensen, and
  Yeager]{1984Oddershede-PolProp}
J.~Oddershede, P.~Jørgensen and D.~L. Yeager, \emph{Computer Physics Reports},
  1984, \textbf{2}, 33 -- 92\relax
\mciteBstWouldAddEndPuncttrue
\mciteSetBstMidEndSepPunct{\mcitedefaultmidpunct}
{\mcitedefaultendpunct}{\mcitedefaultseppunct}\relax
\EndOfBibitem
\bibitem[Giacovazzo \emph{et~al.}(1992)Giacovazzo, Monaco, Viterbo, Scordari,
  Gilli, Zanotti, and Catti]{1992Giacovazzo-BOOK}
C.~Giacovazzo, H.~L. Monaco, D.~Viterbo, F.~Scordari, G.~Gilli, G.~Zanotti and
  M.~Catti, \emph{Fundamentals of Crystallography}, International Union of
  Crystallography, 1st edn, 1992\relax
\mciteBstWouldAddEndPuncttrue
\mciteSetBstMidEndSepPunct{\mcitedefaultmidpunct}
{\mcitedefaultendpunct}{\mcitedefaultseppunct}\relax
\EndOfBibitem
\bibitem[Gonze \emph{et~al.}(2020)Gonze, Amadon, Antonius, Arnardi, Baguet,
  Beuken, Bieder, Bottin, Bouchet, Bousquet, Brouwer, Bruneval, Brunin,
  Cavignac, Charraud, Chen, Côté, Cottenier, Denier, Geneste, Ghosez,
  Giantomassi, Gillet, Gingras, Hamann, Hautier, He, Helbig, Holzwarth, Jia,
  Jollet, Lafargue-Dit-Hauret, Lejaeghere, Marques, Martin, Martins, Miranda,
  Naccarato, Persson, Petretto, Planes, Pouillon, Prokhorenko, Ricci,
  Rignanese, Romero, Schmitt, Torrent, van Setten, Troeye, Verstraete, Zérah,
  and Zwanziger]{2020Gonze-Abinit}
X.~Gonze, B.~Amadon, G.~Antonius, F.~Arnardi, L.~Baguet, J.-M. Beuken,
  J.~Bieder, F.~Bottin, J.~Bouchet, E.~Bousquet, N.~Brouwer, F.~Bruneval,
  G.~Brunin, T.~Cavignac, J.-B. Charraud, W.~Chen, M.~Côté, S.~Cottenier,
  J.~Denier, G.~Geneste, P.~Ghosez, M.~Giantomassi, Y.~Gillet, O.~Gingras,
  D.~R. Hamann, G.~Hautier, X.~He, N.~Helbig, N.~Holzwarth, Y.~Jia, F.~Jollet,
  W.~Lafargue-Dit-Hauret, K.~Lejaeghere, M.~A. Marques, A.~Martin, C.~Martins,
  H.~P. Miranda, F.~Naccarato, K.~Persson, G.~Petretto, V.~Planes, Y.~Pouillon,
  S.~Prokhorenko, F.~Ricci, G.-M. Rignanese, A.~H. Romero, M.~M. Schmitt,
  M.~Torrent, M.~J. van Setten, B.~V. Troeye, M.~J. Verstraete, G.~Zérah and
  J.~W. Zwanziger, \emph{Computer Physics Communications}, 2020, \textbf{248},
  107042\relax
\mciteBstWouldAddEndPuncttrue
\mciteSetBstMidEndSepPunct{\mcitedefaultmidpunct}
{\mcitedefaultendpunct}{\mcitedefaultseppunct}\relax
\EndOfBibitem
\bibitem[Perdew(1985)]{1985Perdew-bandgaperror}
J.~P. Perdew, \emph{International Journal of Quantum Chemistry}, 1985,
  \textbf{28}, 497--523\relax
\mciteBstWouldAddEndPuncttrue
\mciteSetBstMidEndSepPunct{\mcitedefaultmidpunct}
{\mcitedefaultendpunct}{\mcitedefaultseppunct}\relax
\EndOfBibitem
\bibitem[Botti \emph{et~al.}(2004)Botti, Sottile, Vast, Olevano, Reining,
  Weissker, Rubio, Onida, Del~Sole, and Godby]{2004Botti-scissor}
S.~Botti, F.~Sottile, N.~Vast, V.~Olevano, L.~Reining, H.-C. Weissker,
  A.~Rubio, G.~Onida, R.~Del~Sole and R.~W. Godby, \emph{Phys. Rev. B}, 2004,
  \textbf{69}, 155112\relax
\mciteBstWouldAddEndPuncttrue
\mciteSetBstMidEndSepPunct{\mcitedefaultmidpunct}
{\mcitedefaultendpunct}{\mcitedefaultseppunct}\relax
\EndOfBibitem
\bibitem[Benedict \emph{et~al.}(1998)Benedict, Shirley, and
  Bohn]{1998Benedict-dielectric}
L.~X. Benedict, E.~L. Shirley and R.~B. Bohn, \emph{Phys. Rev. B}, 1998,
  \textbf{57}, R9385--R9387\relax
\mciteBstWouldAddEndPuncttrue
\mciteSetBstMidEndSepPunct{\mcitedefaultmidpunct}
{\mcitedefaultendpunct}{\mcitedefaultseppunct}\relax
\EndOfBibitem
\bibitem[Kubo(1966)]{1966Kubo-FD_theorem}
R.~Kubo, \emph{Reports on Progress in Physics}, 1966, \textbf{29},
  255--284\relax
\mciteBstWouldAddEndPuncttrue
\mciteSetBstMidEndSepPunct{\mcitedefaultmidpunct}
{\mcitedefaultendpunct}{\mcitedefaultseppunct}\relax
\EndOfBibitem
\bibitem[Landau and Lifshitz(1980)]{1980Landau-BOOK9.2}
L.~D. Landau and E.~M. Lifshitz, \emph{Course of Theoretical Physics}, Pergamon
  Press, Oxford, 1980, vol.~9\relax
\mciteBstWouldAddEndPuncttrue
\mciteSetBstMidEndSepPunct{\mcitedefaultmidpunct}
{\mcitedefaultendpunct}{\mcitedefaultseppunct}\relax
\EndOfBibitem
\bibitem[Barnett \emph{et~al.}(1992)Barnett, Huttner, and
  Loudon]{1992Barnett-SpontaneousEmissionDielectric}
S.~M. Barnett, B.~Huttner and R.~Loudon, \emph{Phys. Rev. Lett.}, 1992,
  \textbf{68}, 3698--3701\relax
\mciteBstWouldAddEndPuncttrue
\mciteSetBstMidEndSepPunct{\mcitedefaultmidpunct}
{\mcitedefaultendpunct}{\mcitedefaultseppunct}\relax
\EndOfBibitem
\bibitem[Klyshko(1988)]{1988Klyshko-BOOK}
D.~N. Klyshko, \emph{Photons and Nonlinear Optics}, CRC Press, 1988\relax
\mciteBstWouldAddEndPuncttrue
\mciteSetBstMidEndSepPunct{\mcitedefaultmidpunct}
{\mcitedefaultendpunct}{\mcitedefaultseppunct}\relax
\EndOfBibitem
\bibitem[Singer(2012)]{2012Singer-thesis}
A.~Singer, \emph{PhD thesis}, University of Hamburg, 2012\relax
\mciteBstWouldAddEndPuncttrue
\mciteSetBstMidEndSepPunct{\mcitedefaultmidpunct}
{\mcitedefaultendpunct}{\mcitedefaultseppunct}\relax
\EndOfBibitem
\bibitem[Vartanyants \emph{et~al.}(2011)Vartanyants, Singer, Mancuso, Yefanov,
  Sakdinawat, Liu, Bang, Williams, Cadenazzi, Abbey, Sinn, Attwood, Nugent,
  Weckert, Wang, Zhu, Wu, Graves, Scherz, Turner, Schlotter, Messerschmidt,
  L\"uning, Acremann, Heimann, Mancini, Joshi, Krzywinski, Soufli,
  Fernandez-Perea, Hau-Riege, Peele, Feng, Krupin, Moeller, and
  Wurth]{2011Vartanyants-CohLCLS}
I.~A. Vartanyants, A.~Singer, A.~P. Mancuso, O.~M. Yefanov, A.~Sakdinawat,
  Y.~Liu, E.~Bang, G.~J. Williams, G.~Cadenazzi, B.~Abbey, H.~Sinn, D.~Attwood,
  K.~A. Nugent, E.~Weckert, T.~Wang, D.~Zhu, B.~Wu, C.~Graves, A.~Scherz, J.~J.
  Turner, W.~F. Schlotter, M.~Messerschmidt, J.~L\"uning, Y.~Acremann,
  P.~Heimann, D.~C. Mancini, V.~Joshi, J.~Krzywinski, R.~Soufli,
  M.~Fernandez-Perea, S.~Hau-Riege, A.~G. Peele, Y.~Feng, O.~Krupin, S.~Moeller
  and W.~Wurth, \emph{Phys. Rev. Lett.}, 2011, \textbf{107}, 144801\relax
\mciteBstWouldAddEndPuncttrue
\mciteSetBstMidEndSepPunct{\mcitedefaultmidpunct}
{\mcitedefaultendpunct}{\mcitedefaultseppunct}\relax
\EndOfBibitem
\bibitem[Chubar \emph{et~al.}(2013)Chubar, Fluerasu, Berman, Kaznatcheev, and
  Wiegart]{2013Chubar-SRWWavefront}
O.~Chubar, A.~Fluerasu, L.~Berman, K.~Kaznatcheev and L.~Wiegart, \emph{Journal
  of Physics: Conference Series}, 2013, \textbf{425}, 162001\relax
\mciteBstWouldAddEndPuncttrue
\mciteSetBstMidEndSepPunct{\mcitedefaultmidpunct}
{\mcitedefaultendpunct}{\mcitedefaultseppunct}\relax
\EndOfBibitem
\bibitem[Klementiev and Chernikov(2014)]{2014Klementiev-XRT}
K.~Klementiev and R.~Chernikov, Advances in Computational Methods for X-Ray
  Optics III, 2014, pp. 60 -- 75\relax
\mciteBstWouldAddEndPuncttrue
\mciteSetBstMidEndSepPunct{\mcitedefaultmidpunct}
{\mcitedefaultendpunct}{\mcitedefaultseppunct}\relax
\EndOfBibitem
\bibitem[Samoylova \emph{et~al.}(2016)Samoylova, Buzmakov, Chubar, and
  Sinn]{2016Samoylova-WPG}
L.~Samoylova, A.~Buzmakov, O.~Chubar and H.~Sinn, \emph{Journal of Applied
  Crystallography}, 2016, \textbf{49}, 1347--1355\relax
\mciteBstWouldAddEndPuncttrue
\mciteSetBstMidEndSepPunct{\mcitedefaultmidpunct}
{\mcitedefaultendpunct}{\mcitedefaultseppunct}\relax
\EndOfBibitem
\bibitem[Kleinman(1968)]{1968Kleinman-TheoryOPN}
D.~A. Kleinman, \emph{Phys. Rev.}, 1968, \textbf{174}, 1027--1041\relax
\mciteBstWouldAddEndPuncttrue
\mciteSetBstMidEndSepPunct{\mcitedefaultmidpunct}
{\mcitedefaultendpunct}{\mcitedefaultseppunct}\relax
\EndOfBibitem
\bibitem[Freund and Levine(1969)]{1969FreundLevine-parametricconv}
I.~Freund and B.~F. Levine, \emph{Phys. Rev. Lett.}, 1969, \textbf{23},
  854--857\relax
\mciteBstWouldAddEndPuncttrue
\mciteSetBstMidEndSepPunct{\mcitedefaultmidpunct}
{\mcitedefaultendpunct}{\mcitedefaultseppunct}\relax
\EndOfBibitem
\bibitem[Rebuffi and del Rio(2017)]{rebuffi2017oasys}
L.~Rebuffi and M.~S. del Rio, Advances in Computational Methods for X-Ray
  Optics IV, 2017, p. 103880S\relax
\mciteBstWouldAddEndPuncttrue
\mciteSetBstMidEndSepPunct{\mcitedefaultmidpunct}
{\mcitedefaultendpunct}{\mcitedefaultseppunct}\relax
\EndOfBibitem
\bibitem[Pennicard \emph{et~al.}(2013)Pennicard, Lange, Smoljanin, Hirsemann,
  Graafsma, Epple, Zuvic, Lampert, Fritzsch, and
  Rothermund]{pennicard2013lambda}
D.~Pennicard, S.~Lange, S.~Smoljanin, H.~Hirsemann, H.~Graafsma, M.~Epple,
  M.~Zuvic, M.~Lampert, T.~Fritzsch and M.~Rothermund, Journal of Physics:
  Conference Series, 2013, p. 062010\relax
\mciteBstWouldAddEndPuncttrue
\mciteSetBstMidEndSepPunct{\mcitedefaultmidpunct}
{\mcitedefaultendpunct}{\mcitedefaultseppunct}\relax
\EndOfBibitem
\bibitem[Stepanov(2004)]{stepanov2004x}
S.~A. Stepanov, Advances in Computational Methods for X-ray and Neutron Optics,
  2004, pp. 16--26\relax
\mciteBstWouldAddEndPuncttrue
\mciteSetBstMidEndSepPunct{\mcitedefaultmidpunct}
{\mcitedefaultendpunct}{\mcitedefaultseppunct}\relax
\EndOfBibitem
\bibitem[Mikhalychev \emph{et~al.}(2015)Mikhalychev, Benediktovitch,
  Ulyanenkova, and Ulyanenkov]{mikhalychev2015ab}
A.~Mikhalychev, A.~Benediktovitch, T.~Ulyanenkova and A.~Ulyanenkov,
  \emph{Journal of applied crystallography}, 2015, \textbf{48}, 679--689\relax
\mciteBstWouldAddEndPuncttrue
\mciteSetBstMidEndSepPunct{\mcitedefaultmidpunct}
{\mcitedefaultendpunct}{\mcitedefaultseppunct}\relax
\EndOfBibitem
\bibitem[Neumann \emph{et~al.}(1994)Neumann, R{\"u}tt, Bouchard, Schneider, and
  Nagasawa]{neumann1994resolution}
H.-B. Neumann, U.~R{\"u}tt, R.~Bouchard, J.~Schneider and H.~Nagasawa,
  \emph{Journal of applied crystallography}, 1994, \textbf{27},
  1030--1038\relax
\mciteBstWouldAddEndPuncttrue
\mciteSetBstMidEndSepPunct{\mcitedefaultmidpunct}
{\mcitedefaultendpunct}{\mcitedefaultseppunct}\relax
\EndOfBibitem
\bibitem[R{\"u}tt \emph{et~al.}(1995)R{\"u}tt, Neumann, Poulsen, and
  Schneider]{rutt1995resolution}
U.~R{\"u}tt, H.-B. Neumann, H.~Poulsen and J.~Schneider, \emph{Journal of
  applied crystallography}, 1995, \textbf{28}, 729--737\relax
\mciteBstWouldAddEndPuncttrue
\mciteSetBstMidEndSepPunct{\mcitedefaultmidpunct}
{\mcitedefaultendpunct}{\mcitedefaultseppunct}\relax
\EndOfBibitem
\bibitem[Boemer(2020)]{DissBoemer}
C.~Boemer, \emph{PhD thesis}, Universit\"at Hamburg, 2020\relax
\mciteBstWouldAddEndPuncttrue
\mciteSetBstMidEndSepPunct{\mcitedefaultmidpunct}
{\mcitedefaultendpunct}{\mcitedefaultseppunct}\relax
\EndOfBibitem
\bibitem[Glover \emph{et~al.}(2012)Glover, Fritz, Cammarata, Allison, Coh,
  Feldkamp, Lemke, Zhu, Feng, Coffee,\emph{et~al.}]{glover2012x}
T.~Glover, D.~Fritz, M.~Cammarata, T.~Allison, S.~Coh, J.~Feldkamp, H.~Lemke,
  D.~Zhu, Y.~Feng, R.~Coffee \emph{et~al.}, \emph{Nature}, 2012, \textbf{488},
  603--608\relax
\mciteBstWouldAddEndPuncttrue
\mciteSetBstMidEndSepPunct{\mcitedefaultmidpunct}
{\mcitedefaultendpunct}{\mcitedefaultseppunct}\relax
\EndOfBibitem
\bibitem[Popova-Gorelova \emph{et~al.}(2018)Popova-Gorelova, Reis, and
  Santra]{popova2018theory}
D.~Popova-Gorelova, D.~A. Reis and R.~Santra, \emph{Physical Review B}, 2018,
  \textbf{98}, 224302\relax
\mciteBstWouldAddEndPuncttrue
\mciteSetBstMidEndSepPunct{\mcitedefaultmidpunct}
{\mcitedefaultendpunct}{\mcitedefaultseppunct}\relax
\EndOfBibitem
\end{mcitethebibliography}
%\bibliographystyle{unsrt} %our initial choice
%\bibliography{rsc} %You need to replace "rsc" on this line with the name of your .bib file
\bibliographystyle{rsc} %the RSC's .bst file

\end{document}